\documentclass[prb,twocolumn,a4paper,superscriptaddress,floatfix,showpacs,amsmath,amssymb]{revtex4}

\usepackage{amssymb} \usepackage{amsmath} \usepackage{amsfonts}
\usepackage{graphicx} \usepackage{bm} \usepackage{color}

\usepackage{subfig}

\usepackage{listings} \usepackage{hyphenat}
\usepackage[cp1251]{inputenc}
 \usepackage[T2A]{fontenc}
 \usepackage[english]{babel}
\usepackage{color}
\usepackage{amsthm}
\usepackage{braket}

\usepackage{verbatim}

\usepackage{yfonts}[1998/10/03] \usepackage{mathrsfs}

\renewcommand{\Im}{\,\textrm{Im}\,} 
 \DeclareMathOperator{\tr}{tr}

 1 

\DeclareMathAlphabet{\mathscrbf}{OMS}{mdugm}{b}{n}

\renewcommand{\Im}{\mathop\mathrm{Im}\nolimits}

 1 
scaled \magstep 1

 \def\ve{\varepsilon} 
\def\vf{\varphi}

 \def\de{\partial} 
\def\tr{\hbox{\,tr\,}} 
 \def\br{\mathbf{r}} \def\bp{\mathbf{p}}
\def\bq{\mathbf{q}} 
  \def\br{\mathbf{r}}
\def\brp{\mathbf{r}^\prime} 
\def\bsigma{\bm{\sigma}} 
 \def\brp{\mathbf{r}^\prime}
 
  \def\bn{\mathbf{n}}

\def\pf{{p_{\textrm{F}}}} \def\ef{\varepsilon_{\textrm{F}}}
\def\pe{{p_{\varepsilon}}} \def\ef{\varepsilon_{\textrm{F}}}

\begin{document}

\title{Effects of anisotropy and disorder on the conductivity of Weyl semimetals}

\author{Ya.~I.~Rodionov} \affiliation{Institute for Theoretical
and Applied Electrodynamics, Russian Academy of Sciences,
Izhorskaya Str. 13, Moscow, 125412 Russia}

\author{K.~I.~Kugel} \affiliation{Institute for Theoretical and
Applied Electrodynamics, Russian Academy of Sciences, Izhorskaya
Str. 13, Moscow, 125412 Russia} \affiliation{Center for Emergent
Matter Science, RIKEN, Saitama, 351-0198, Japan}

\author{Franco~Nori} \affiliation{Center for Emergent Matter
Science, RIKEN, Saitama, 351-0198, Japan} \affiliation{Department
of Physics, University of Michigan, Ann Arbor, MI 48109-1040, USA}

\begin{abstract} We study dc conductivity of a Weyl semimetal with uniaxial anisotropy (Fermi velocity ratio $\xi= v_\bot/v_\parallel\neq1$)  considering the scattering of charge carriers by a wide class of impurity potentials, both short- and long-range. We obtain the ratio of transverse and longitudinal (with respect to the anisotropy axis) conductivities as a function of both $\xi$ and temperature. We find that the transverse and longitudinal conductivities exhibit different temperature dependence in the case of short-range disorder. For general long-range disorder, the temperature dependence ($\sim T^4$) of the conductivity turns out to be insensitive of the anisotropy in the limits of strong ($\xi\gg$ and $\ll1$) and weak ($\xi\approx1$) anisotropy.
\end{abstract}

\date{\today}

\maketitle


\section{Introduction\label{Sec:Intro}}

Weyl semimetals (WSM) are three-dimensional (3D) analogs of graphene.~\cite{Vafek_AnnRevCondMat2014,Turner_arxiv2013} Their quasiparticles are described by the massless 3D Dirac Hamiltonian. Such systems were first proposed as an exotic theoretical possibility, with an expectation to observe some of their features in pyrochlore iridates~\cite{Wan_PRB2011} and in certain semiconductor heterostructures.~\cite{Burkov_PRL2011,HalaszPRB2012}
Recent experiments uncovered several chemical compounds, which can be classified as WSM~\cite{Neupane_NatCom2013,Liu_Science2014,Liu_NatMat2014,
Jeon_NatMat2014,Borisenko_PRL2014,LuArXiv:15,XuArXiv:15,LvArXiv:15}.
The subject is attracting growing theoretical and experimental interest. (see, e.g., the reviews in
Refs.~\onlinecite{Vafek_AnnRevCondMat2014,Turner_arxiv2013,Hosur_ComptRend2013}).

Charge transport in WSM has also received considerable attention and a number of interesting phenomena have been discovered (see Ref.~\onlinecite{Hosur_ComptRend2013} and references therein). The main emphasis was on manifestations of topological effects: an additional topological protection of the gapless spectrum near the Dirac points in 3D and the chiral anomaly,~\cite{HosurPRL2012,SonPRB2013,ChenPRB2013,BiswasPRB2014,PanfilovPRB2014,
ParameswaranPRX2014,WitczakPRL2014} as well as an unusual Kondo effect~\cite{Principi_arxiv2014}. Disorder and impurity effects have also attracted significant attention.~\cite{HuangPRB2013,HuangNJP013,OminatoPRB2014,RodioSyzr_PRB2015}
A detailed study of the influence of Coulomb disorder in highly compensated WSM was undertaken in Ref.~\onlinecite{SkinnerPRB2014}. The problem is that even in the simplest case of delta-correlated disorder, the field theory of a WSM is non-renormalizable. Progress in overcoming this stumbling block was recently achieved in
Refs.~\onlinecite{Syzranov_arxiv2014,Syzranov_PRB2015}, where the conductivity of weakly-disordered semimetals was treated in terms of the $\ve$-expansion within the renormalization group (RG) approach. However, Weyl semimetals correspond to $\ve=-1$ and the predictions are qualitative. Nevertheless, it appeared to be possible to reveal some specific features of the conductivity for different kinds of disorder and even to predict a disorder-driven quantum phase transition distinct from the Anderson transition. A RG approach~\cite{Moon_arxiv2014} has been also applied to demonstrate the possibility of non-Fermi-liquid behavior in  systems with isotropic 3D Dirac points having a weak disorder and Coulomb-type interparticle interactions.~\cite{Moon_arxiv2014}

The computation of the isotropic conductivity was recently performed~\cite{OminatoPRB2014,SarmaPRB2015,Rama} in the framework of the kinetic-equation technique, improved by the self-consistent Born approximation, as well as numerically.~\cite{Bjorn}

Most Weyl semimetals obtained in laboratories so far are anisotropic.~\cite{Liu_Science2014,Liu_NatMat2014} The anisotropy can be also induced by a linearly-polarized electromagnetic wave.~\cite{Syzr-Rod2013} Some effects related to the anisotropy and tilting of the Dirac cones have recently been treated.~\cite{Bjorn_PRB2015} However, the influence of anisotropy on the charge transport in WSM with quenched disorder has remained unexplored so far. In this paper, we analyze the effects of disorder and anisotropy on the conductivity, in the framework of the Born approximation. We treat these effects in the diagrammatic approach, both for short- and long-range disorder taking into account the uniaxial anisotropy. We obtain analytical results for the cases of strong and weak anisotropy. The diagrammatic framework allows us to find out the validity range of the Fermi-liquid approach in WSM.

The potential produced by impurities (disorder potential) is introduced as a rather general spatial distribution $u(\br)$. Its characteristic scale in momentum space is specified by a parameter $p_0$. Thus, at small values of $p_0$, we are dealing with long-range disorder, whereas large $p_0$ corresponds to short-range disorder. The ratio of Fermi velocities perpendicular and parallel to the $z$ axis, $\xi=v_\bot/v_\parallel$, becomes an additional control parameter of the problem.

We have found that for sufficiently low temperatures (or doping level $\ef$), when the disorder can be considered to be a short-range one ($p_0\gg\max\{T/v_\parallel,\ef/v\}$), the longitudinal and transverse conductivities exhibit a different temperature dependence
\begin{gather}
   \label{res-1}
     \frac{\sigma_\parallel\xi^2}{\sigma_\bot}=
     1-\frac{4\de g(0)}{15}(1+\xi^{-2})\frac{\pi^2 T^2/3+\ef^2}{(v_\parallel p_0)^2}\,,
  \end{gather}
where $g(p^2/p_0^2)$ is the impurity structure factor and $\de g(0)\equiv dg(x)/dx|_{x=0}$. The temperature dependence in Eq.~\eqref{res-1} is in fact the first term of the asymptotic series in $\max\{T/v_\parallel,\ef/v\}/p_0\ll1$. This dependence becomes even more pronounced at higher temperatures $T\sim p_0 v_\parallel$.

The $\sigma_\parallel\xi^2/\sigma_\bot$ ratio saturates to a temperature-independent constant in the high-temperature (or long-range disorder) limit, when ($p_0\ll\max\{T/v_\parallel,\ef/v\}$). At $\ef\ll T$, the temperature dependence of both components of the conductivity obeys the relation
\begin{gather}
   \sigma_{\parallel},\, \sigma_\perp\, \sim \, T^4
\end{gather}
regardless of the particular form of the disorder potential or impurity structure factor. In the long-range disorder limit, for  $\xi\gg1,\ \xi\approx1,$ and $\xi\ll1$, we found that the
$\sigma_\parallel\xi^2/\sigma_\bot$ ratio approaches constant values, independent of temperature  and disorder potential. The results concerning the conductivity ratio can be summarized in the form
\begin{gather}
   \frac{\sigma_\parallel\xi^2}{\sigma_\bot}=
      \begin{cases}
         3/\left(\frac{1}{2}+4\ln 2 \right)+\mathcal{O}(\xi^{-1}),\ &\ \xi\gg1,\\
         1+\mathcal{O}(\delta\xi^{2}),\ \ \xi=1+\delta\xi,\ &\ \delta\xi\ll1,\\
            c_3+\mathcal{O}(\xi^2),\ &\ \xi\ll1,
      \end{cases}
\end{gather}
where $c_3$ is a constant of the order of unity.

The scales of the problem are the temperature $T$ of the system, Fermi energy $\ef$, and the elastic scattering rate $1/\tau$. The latter is assumed to be small, so
\begin{gather}
\frac{1}{\max\{\ef,T\}\tau}\ll1
\end{gather}
is a small parameter of the problem. Our task is to compute the Drude conductivity. The fermion doubling theorem implies that the WSM spectrum always has an even number of Weyl points. For simplicity, we assume that the charge carriers have identical spectra near all Weyl points and compute the conductivity \textit{per point}.  All the results for the conductivity should be multiplied by the number of Weyl points.

The disorder potential is assumed to be quite weak, so that the Born approximation is justified
\begin{gather}
   |u(\br)|\ll \min\{v_\bot,v_\parallel\}p_0\, .
\end{gather}
Note here that our results are applicable not only to the Weyl semimetals themselves, but to a wider class of materials with the 3D Dirac spectrum (topological Dirac semimetals etc., see the classification given in Ref.~\onlinecite{Nagaosa_NatCom2014}).

In Section II, we describe the formalism used in dealing with the scattering problem, give all the necessary definitions, and describe the employed calculation procedures. In this section, we also analyze the limits of applicability of the Fermi-liquid approach in the case of WSM. In Section III, we revise the general temperature dependence of the conductivity in isotropic Weyl semimetals with long- and short-range disorder. Section IV deals with the  anisotropic case. We provide a detailed analysis of the solution to the Dyson equation for the singular part of the vertex function for short- and long-range disorder potentials. We compute the conductivity for different $T/\ef$ ratios and different values of the anisotropy parameter $\xi$. The conclusions are presented in Section V. Several important technical issues are presented in the Appendices.

\section{Formalism\label{Sec:I}}

\subsection{The model}

We consider a system with a Hamiltonian, which generally has uniaxial anisotropy. Namely, the Fermi velocity along a specified axis $\bn_0$ (the longitudinal component) is different from that in the perpendicular direction $v_{\perp}=\xi v_\parallel$
\begin{gather}
  \label{h1}
 \begin{split}
  H&=H_0+H_{\textrm{dis}};\\
  H_0&=-i\int\psi^\dag(\br)\big[v_\parallel \sigma_\parallel
  \de_{r_\parallel}+v_\bot \bsigma_\bot\de_{\br_\bot}\big]\psi(\br) d\br\, ,\\
   H_{\textrm{dis}}&=\int\psi^\dag(\br)u(\br)\psi(\br) d\br\,,
 \end{split}
\end{gather}
where $u(\br)$ is the disorder potential, $\psi$
are quasiparticle field operators, and $\sigma_\parallel=\bsigma\bn_0,\ r_\parallel=\br\bn_0$,
$\bsigma_\bot=\bsigma-\bsigma\bn_0,\ \br_\bot=\br-\bn_0(\bn_0\br)$ are the projections of Pauli matrices and radius vectors.
The disorder potential correlation function reads
\begin{gather}
    \label{form-factor}
    \int d\br e^{-i\bp\br}\langle u(\br)u(0)\rangle =\frac{n_{\text{imp}}u_0^2}{p^6_0}\,g\left(\frac{p^2}{p_0^2}\right)\, ,
\end{gather}
where $g(p^2/p^2_0)$ is the dimensionless Fourier transform of the normalized ($g(0)=1$) disorder structure factor.
In our case, it incorporates the correlation of impurity positions, as well as the form of the potential. The parameter $u_0$ plays the role of the amplitude of the disorder potential and $n_{\text{imp}}$ is the concentration of impurities. The potential is assumed to be isotropic. The last assumption certainly depends on the nature of the disorder. For example, the isotropic Coulomb potential may acquire an anisotropic screening by charged carriers with an anisotropic spectrum. Here, we neglect the induced anisotropy of the potential. Our task is to emphasize the main effect of the anisotropic spectrum on the transport properties, which manifests itself even with an isotropic potential.

To simplify the analysis, we rescale the coordinates and $\psi$
operators to absorb the anisotropy in $H_0$ according to
\begin{gather}
  \label{rescale}
 \begin{split}
  r_\parallel=r^\prime_\parallel,\ \
  \br_\bot=\xi r^\prime_\bot,\ \
   \psi(\br)=\frac{1}{\xi}
   \psi^\prime(\br^\prime).
 \end{split}
 \end{gather}
 Then, the Hamiltonian changes its form to
 \begin{gather}
  \label{ham-renorm}
  \begin{split}
  H^\prime_0&=-iv_\parallel\int\psi^{\prime\dag}(\brp)
  (\bsigma\de_\brp)\psi^{\prime}(\brp)d\brp,\\
  H_{\textrm{dis}}^\prime&=\int d\br^\prime \psi^{\prime\dag}(\brp)
  u\big(\brp_\parallel+\xi\brp_\bot\big)
  \psi^{\prime}(\brp)\, ,
  \end{split}
 \end{gather}
which leads to the modified disorder correlation function
 \begin{gather}
   \int d\br^\prime e^{-i\bp\br^\prime}
    \left\langle u\big(\brp_\parallel+\xi\brp_\bot\big)u(0)\right\rangle=
\frac{n_{\text{imp}}u_0^2}{p^6_0\xi^2}\,g^\prime
\left(\frac{p^2}{p_0^2}\right),\notag\\
   g^\prime\left(\frac{p^2}{p_0^2}\right) =g
    \left(\frac{(1-\xi^{-2})(\bp\bn_0)^2
     +\xi^{-2}p^2}{p_0^2}\right),
     \label{disord1}
\end{gather}
where we have decomposed the momentum as: $p^2=p_\parallel^2+(p^2-p_\parallel^2)\rightarrow ({\bp\bn_0})^2+\xi^{-2}[p^2-({\bp\bn_0})^2]$.

\begin{figure}[h]
\begin{center}
  \includegraphics[width=0.8\columnwidth]{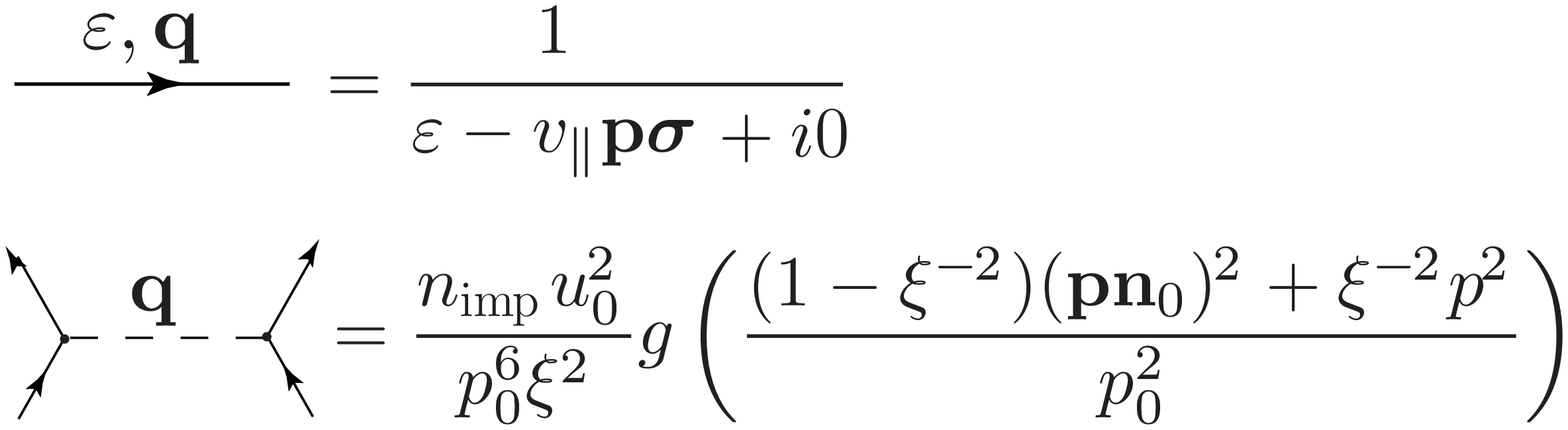}
\end{center}
   \caption{\label{feyn}
Feynman diagrams for the non-interacting fermion retarded Green's function and the disorder correlation function.
  }
\end{figure}

The Feynman rules are extracted from~\eqref{ham-renorm} and~\eqref{disord1}, and depicted in Fig.~\ref{feyn}.

\subsection{Conductivity}
The conductivity tensor is found via the Kubo formula
\begin{gather}
  \label{kubo1}
  \begin{split}
   \sigma_{\alpha\beta}(\omega,0)&=\frac{e^2\Pi^R_{\alpha\beta}
   (\omega)}{i\omega}\, ,\\
   \Pi^R_{\alpha\beta}(\omega)&=i\int dt\,d\br
   \langle[j_\alpha(t,\br),j_\beta(0,0)]\rangle
   e^{i\omega t}\theta(t)\, ,
 \end{split}
\end{gather}
where $\Pi^R_{\alpha\beta}(\omega)$ is the retarded polarization operator and $j_\alpha(t,\br)=\psi^\dag(t,\br)\sigma_\alpha v_\alpha\psi(t,\br)$ is the quasiparticle current operator. The averaging $\langle...\rangle$ is assumed to be done over the Gibbs distribution as well as over different realizations of disorder. We will be interested in the system response to a uniform  electric field, constant in time. For this purpose, we set $\bq=0$ and $\omega\rightarrow0$.

\begin{figure}[h] \centering
  \includegraphics[width=0.85\columnwidth]{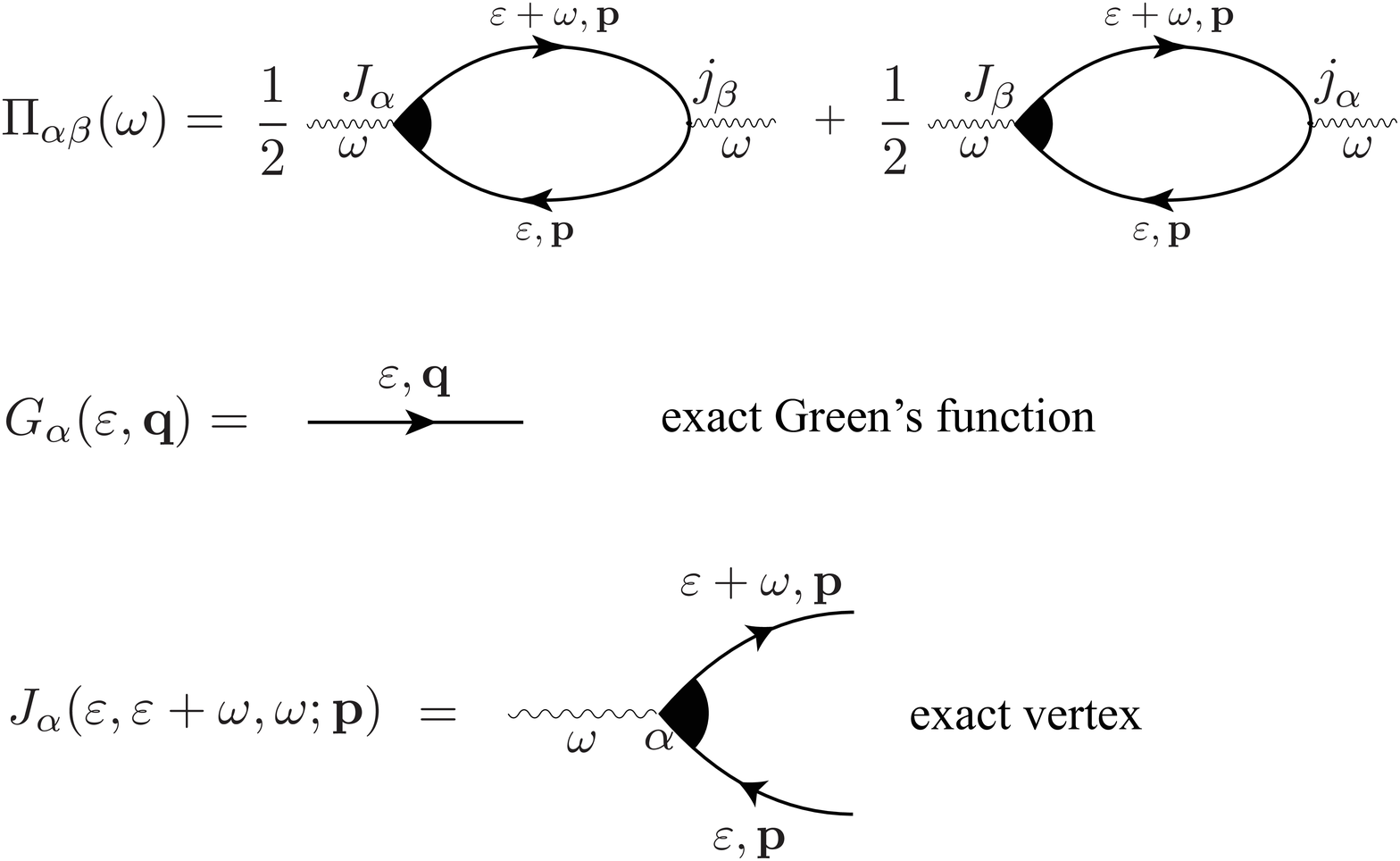}
   \caption{\label{diag1} Diagrammatic representation of the polarization operator.}
 \end{figure}

The computation of the polarization operator in the lowest order of the $1/(\ef\tau)$ expansion involves the summation of disorder ladder series (accounting for the difference between averages $\langle j_{\alpha}j_\beta\rangle$ and $\langle j_{\alpha}\rangle\langle j_\beta\rangle$) as well as one-loop corrections to the Green's functions (mainly responsible for the finite quasiparticle lifetime).

Due to the Onsager relations, the conductivity (as well as the polarization operator) is a symmetric tensor. Thus, the exact diagrammatic representation for $\Pi_{\alpha\beta}(\omega)$ (we denoted $\Pi_{\alpha\beta}(\omega,0)\equiv \Pi_{\alpha\beta}(\omega)$) allows for its symmetric form (see
Fig.~\ref{diag1}).

\begin{figure}[t]
\centering
  \includegraphics[width=0.85\columnwidth]{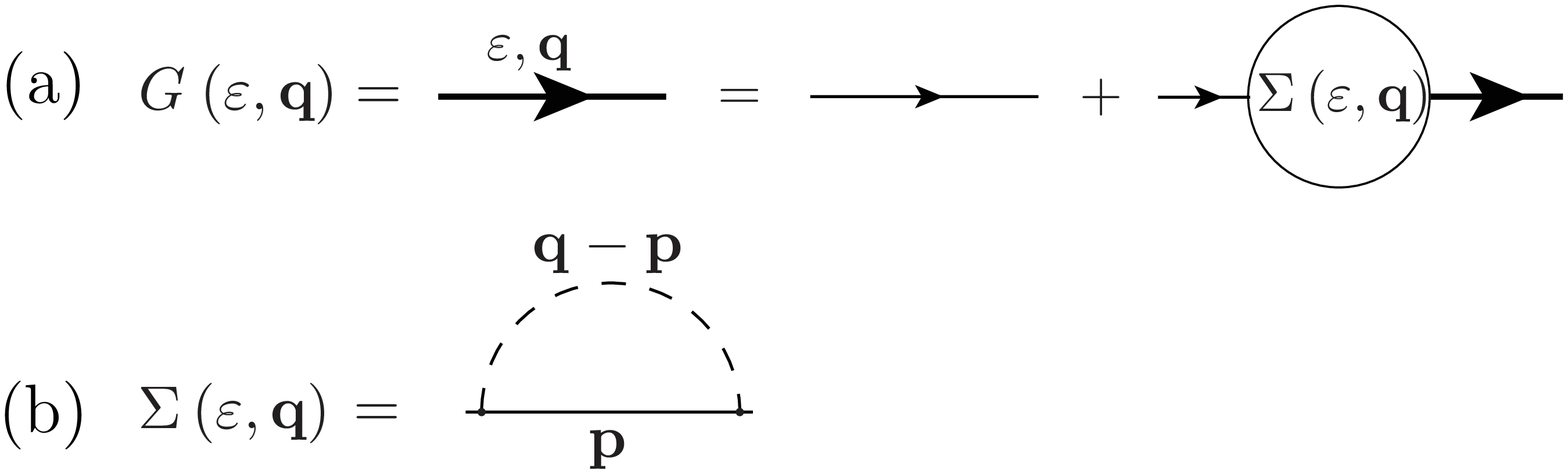}
   \caption{\label{diag0}
(a) Dyson equation for the exact fermion Green's function; (b) Fermion self-energy in the one-loop approximation.
          }
\end{figure}

The non-interacting and disorder-averaged Green's functions are related via the standard diagrammatic equation depicted in
Fig.~\ref{diag0}a and have the form
\begin{gather}
 \begin{split}
 G^R(\ve,\bq)=\big[\ve-v_\parallel\bp\bsigma-\Sigma^R(\ve,\bq)\big]^{-1}.
 \end{split}
 \end{gather}
The
self-energy is worked out in the Born approximation; its diagrammatic representation is shown in Fig.~\ref{diag0}b.

\begin{figure}[t]
\centering
  \includegraphics[width=0.85\columnwidth]{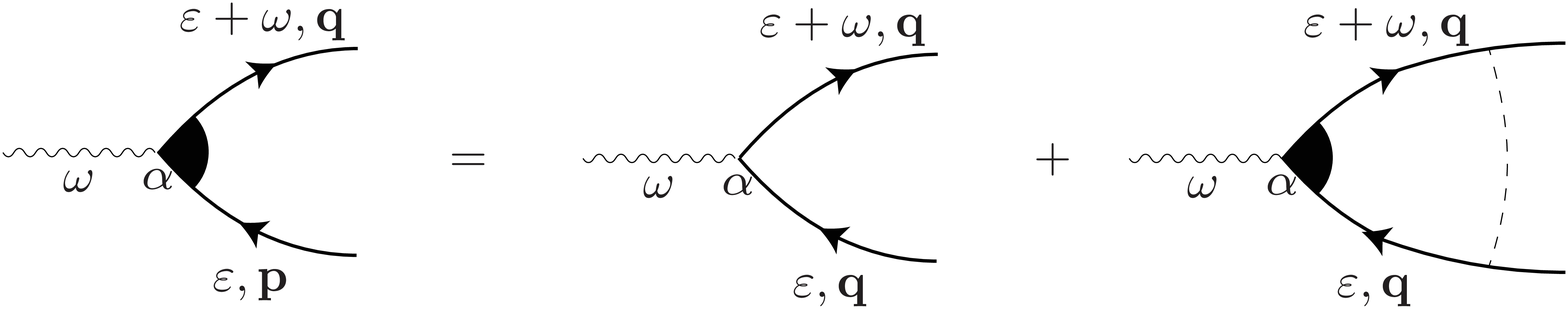}
   \caption{\label{diag2} Dyson equation for the vertex $J_\alpha(\ve,\ve+\omega,\omega;\bp)$.
          }
\end{figure}

The ladder summation in the polarization operator is included into the renormalized vertex $J$ corresponding to the electric current. To perform the summation, one has to solve the corresponding Dyson equation depicted in Fig.~\ref{diag2}.

The conductivity tensor in a system with the uniaxial anisotropy is
characterized by only two eignenvalues $\sigma_\parallel,\ \sigma_\bot$ and in ``principal rescaled axes'' takes the form
\begin{gather}
   \sigma_{\alpha\beta}=\begin{pmatrix}
                       \xi^{2}\sigma_{\parallel} & 0 & 0\\
                       0 & \sigma_{\bot} & 0\\
                       0 & 0 & \sigma_{\bot}
                       \end{pmatrix}\, .
\end{gather}
The general expression for the conductivity in the leading $1/(\ef\tau)$ approximation reads
\begin{gather}
   \label{cond-temp}
   \sigma_{\alpha\beta}(T)=\frac{1}{4T}\int\frac{d\ve}{2\pi}
   \frac{\sigma_{\alpha\beta}(\ve)}{\cosh^2\frac{\ve-\ef}{2T}}\,, \end{gather}
where
\begin{widetext}
\begin{gather}
   \label{polar-fin}
\begin{split}
   \sigma_{\alpha\beta}(\ve)=
   \frac{e^2v_\parallel^2}{4\pi}\int\frac{d\bp}{(2\pi)^3}
   \hbox{tr}\Big\{
   G^A(\ve,\bp)\big[J^{ARR}_\alpha(\ve,
   \bp)-\sigma_\alpha\big]G^R(\ve,\bp)\sigma_\beta+
   G^A(\ve)\big[J^{ARR}_\beta(\ve,\bp)-\sigma_\beta\big]
   G^R(\ve,\bp)\sigma_\alpha\\
   -\big[G^R(\ve,\bp)-G^A(\ve,
   \bp)\big]\sigma_\alpha\big[G^R(\ve,
   \bp)-G^A(\ve,\bp)\big]
   \sigma_\beta
   \Big\}\,.
 \end{split}
\end{gather}
\end{widetext}
Here, for brevity, we denoted $J_\alpha^{ARR}(\ve,\ve,0;\bp)\equiv
J_\alpha^{ARR}(\ve,\bp)$, and $J^{ARR}$ is the singular part of the vertex function (see Appendix A for details).

\subsection{Validity of the Fermi-liquid approach}

Expression~\eqref{polar-fin} for $\sigma_{\alpha\beta}(\ve)$  is rather complicated and deserves special attention. The integrand defining $\sigma_{\alpha\beta}(\ve)$ is deliberately split into two lines. The second line leads to a convergent integral because $(G^R-G^A) \sim 1/p^2$,  when $p\rightarrow\infty$. The first line of~\eqref{polar-fin}though, contains the vertex $J^{ARR}$, which makes it problematic even in the isotropic case ($v_\parallel=v_\bot=v$). At first glance, the expression seems suitable enough, because the product $G^R G^A$ is sharply peaked at $p=p_{\ve}=\ve/v$.  The integral therefore, appears to be completely determined by the vicinity of $p=p_\ve$. However, a more attentive look reveals that the product $G^R G^A$ converges rather slowly: $G^RG^A\sim 1/p^2$ at $\pe \ll p\ll \tau\ve \pe$, and $G^R G^A\sim 1/p^4$ at $p\gg \tau\ve \pe$. Thus, we need to know at least the asymptotic behavior of $J^{ARR}(\ve,\bp)$ at large momentum in order to complete the calculation.

This difficulty is closely related to the justification of the Fermi-liquid approach. Indeed, this justification includes two constraints: (a) the smallness of the quasiparticle scattering rate guarantees  the applicability of perturbation theory (in most cases, the vacuum state, the Fermi sphere, survives); and (b) all observables are determined by the scattering processes in the proximity of the Fermi sphere. The latter statement allows expanding the momentum near the Fermi surface in every integration ($d^3 p=(\ef/v)^2 dp$  in 3D) making any momentum integral one-dimensional and fast converging.

\begin{figure}[b]
\centering
  \includegraphics[width=0.8\columnwidth]{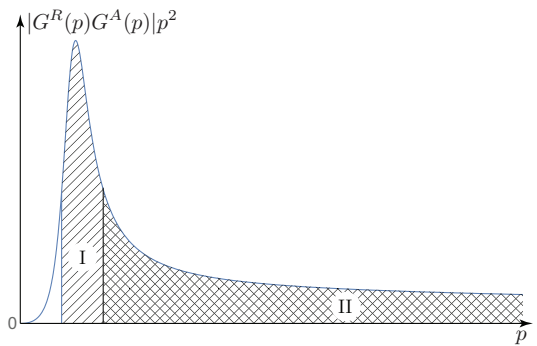}
   \caption{\label{fermi1}
     Contributions to the conductivity. Region I represents the Fermi liquid contribution; region II originates from the scattering far from the Fermi surface.
          }
\end{figure}

The justification of the second constraint follows from the fact that a momentum integral defining the observable always involves the $G^R(p)G^A(p)$ term (sharply peaked near the Fermi sphere). However, we have just seen that the latter argument fails when computing the WSM conductivity. The integrand defining the conductivity does contain a sharp $G^R(p)G^A(p)$ peak \textit{as well as} a long tail (see Fig.~\ref{fermi1}).

%
To justify the Fermi-liquid framework, we need to accurately estimate the contribution given by the tail. We divide the integration domain according to $p^2 dp=(p^2-\pe^2)dp+\pe^2 dp$. Then, $F^{\textrm{sing}}$ is split into two parts
 \begin{gather}
  \label{est1}
  \begin{split}
   &F^{\textrm{sing}}_{\alpha\beta}=\hbox{I}+\hbox{II}\,,\\
   &\hbox{I}=\int (p^2-\pe^2) G^{A}(\ve,\bp)
   [J_\alpha^{ARR}(\ve,\bp)-\sigma_\alpha]G^{R}
   (\ve,\bp)\sigma_\beta dp,\\
   &\hbox{II}=\pe^2\int G^{A}(\ve,\bp)
   [J_\alpha^{ARR}(\ve,\bp)-\sigma_\alpha]G^{R}
   (\ve,\bp)\sigma_\beta\,dp\, .
  \end{split}
\end{gather}
Part II is already convergent. The convergence of part I entirely depends on the asymptotic behavior of the vertex function. The  ultraviolet behavior of the vertex is discussed in Appendix C, where it is proved to decay fast enough to secure its convergence. Part I is estimated in Appendix D, where it is shown that
 \begin{gather}
  \label{comp}
  \hbox{I}\sim\max\left\{1,\frac{p_0}{p_\ve}\right\}\frac{1}{\ve\tau}\hbox{II}\, .
 \end{gather}
This means that the Fermi-liquid approach holds with $1/(\max\{\ef,T\}\tau)$ accuracy. Therefore, some care needs to be taken when working out the weak-localization corrections (WLC) to conductivity in, e.g., the 3D case. Indeed, this correction is determined by the next term in the $1/(\ef\tau)$ expansion, and the WLC in 3D reads $\delta\sigma_{\textrm{\tiny WLC}}/\sigma\sim1/(\ef\tau)$. Therefore, when computing corrections to the conductivity in the 3D case, one has to take into account the values of the momentum far from the Fermi surface.

As a result, expanding all the momentum integrals near the Fermi surface, we can omit (in the leading order) the $G^R G^R$ and $G^A G^A$ terms in the expression~\eqref{polar-fin} for the conductivity and obtain the standard formula
\begin{gather}
   \label{polar-fin1}
\begin{split}
  \sigma_{\alpha\beta}(\ve)&=
   \frac{e^2v_\parallel^2}{4\pi}\int\frac{d\bp}{(2\pi)^3}
   \hbox{tr}\Big\{
   G^A(\ve,\bp)J^{ARR}_\alpha(\ve,
   \bp)G^R(\ve,\bp)\sigma_\beta\\
   &+
   G^A(\ve)J^{ARR}_\beta(\ve,\bp)G^R(\ve,\bp)\sigma_\alpha
   \Big\}\,.
 \end{split}
\end{gather}


\section{Conductivity in the isotropic case}
In order to understand how to tackle the anisotropic problem, we briefly outline the diagrammatic derivation of the isotropic conductivity. Wherever isotropy is implied, we omit the $\parallel$  and $\bot$ indexes, and write $v_\parallel=v_\bot=v$. In the corresponding limits, we reproduce the earlier results for the conductivity.~\cite{OminatoPRB2014,SarmaPRB2015,RodioSyzr_PRB2015}

\subsection{Disorder averaging}

The self-energy of the Green's function has matrix structure (due to the presence of $\sigma$-matrices) and consists of two parts:
$\Sigma^R=\Sigma^R_{\text{I}}+\Sigma^R_{\text{II}}$.
The first part is responsible for the renormalization of Fermi velocities and the rescaling of the Fermi field operators; the second part is responsible for the non-vanishing relaxation rate. Perturbation theory can be applied only if $|\Sigma_R|\ll\max\{\ef,T\}$. The estimate gives
\begin{gather}
  \begin{split}
   \label{self0}
  \frac{|\Sigma^R_{\text{I}}|}{\ve}&\sim \Big(\frac{vp_0}{\ve}\Big)\frac{1}{\tau\ve}\sim \lambda(p_0)\min\left\{1,\left(\frac{p_0v}{\ve}\right)^2\right\},\\
  \frac{1}{\tau}&\sim \frac{(n_{\text{imp}}u_0^2)\min\{\ve^2,(p_0 v)^2\}}{v^3p^6_0}\,,
  \end{split}
\end{gather}
where $|\Sigma^R|$ is understood as the modulus of any of its matrix component, $\ve\sim\max\{T/v,\ef/v\}$ ($\ef$ can be considered here as the the doping level). The dimensionless coupling constant
\begin{gather}
\lambda(p_0)=\frac{n_{\text{imp}} u_0^2}{v^2 p_0^5}
\end{gather}
represents the strength of the potential. It is defined at the characteristic disorder scale $p_0$.

Clearly, despite the smallness of the Fermi-liquid parameter $1/(\ve\tau)$, the first part of the self energy may lead to a strong renormalization for short-range $(p_0\gg \ve/v)$ disorder, depending on the initial coupling strength $\lambda(p_0)$. This is in contrast to the situation in ordinary metals, where the ultraviolet cut-off $vp_0/\ef\sim 1$ and the smallness of the $1/(\ef\tau)$ parameter guarantees the validity of perturbation theory. In WSM, it is appropriate to consider both limits: short- ($p_0\gg \max\{\pf,T/v\}$) and long-range ($p_0\ll \max\{\pf,T/v\}$)  disorder.

\subsubsection{Renormalization due to short-range disorder}

The analysis of the influence of strong short-range disorder $\lambda(p_0)\lesssim 1)$ was attempted in Ref.~\onlinecite{OminatoPRB2014} via the self-consistent Born approximation (SCBA). The latter however overlooks the diagrams with crossed disorder lines. In WSM, these diagrams happen to be of the same order as the diagrams included in the SCBA. Therefore, the SCBA is an uncontrolled approximation for a WSM. A fully consistent analysis was performed in~Refs.~\onlinecite{Syzranov_PRB2015,Syzranov_arxiv2014} (by means of the $\ve$-expansion). Crucially, it was shown that the disorder operator becomes relevant if the initial coupling exceeds some critical value
\begin{gather}
  \lambda(p_0)>\lambda_*\sim 1.
\end{gather}

For $\lambda(p_0)>\lambda_*$, the disorder coupling grows with decreasing
momentum, which leads to the metal--insulator transition. If the initial disorder strength is small, $\lambda(p_0)\ll\lambda_*$, then the disorder operator is irrelevant. The corresponding running value of  the dimensionless coupling constant is
\begin{gather}
   \lambda(p)=\lambda(p_0)\frac{p}{p_0}\, ,
\end{gather}
where $p\sim\max\{\ef/v,T/v\}$. We will focus on the case $\lambda(p_0)\ll1$ and discard the renormalization of the Fermi velocity and field operators.

\subsubsection{Long-range disorder}
As one can deduce from Eq.~\eqref{self0}, the long-range $(p_0\ll\ve/v)$ disorder  does not present a problem since the corresponding $\Sigma_{\text{I}}$ part is much smaller than the $1/(\ve\tau)$ parameter. Therefore, here we also discard the renormalization of the Fermi velocity and field operators.

\subsection{The general Dyson equation and its solution in the isotropic case}
Surprisingly, even for an anisotropic system, the Green's function preserves its simple isotropic form (in the limiting cases of short-range and long-range disorder, see Appendix F with detailed calculations). The Green's function has the form~\cite{RodioSyzr_PRB2015}
\begin{gather}
\label{green-2}
  G^{R,A}(\ve,\bp)=\frac{\ve+ v_\parallel\bp\bsigma}{\big(\ve\pm\frac{i}{2\tau}\big)^2-
  \big(v_\parallel p\mp\frac{i}{2\tau_1}\big)^2}\, .
\end{gather}

We will need the following scattering times
\begin{gather}
  \label{sc}
  \begin{split}
    \frac{1}{\tau(\bn\bn_0)}&=\frac{n_{\text{imp}}\ve^2}{2\pi v_\parallel^3p_0^6}\int\frac{d\Omega}{4\pi}g^\prime_{\pe(\bn-\bn^\prime),\bn_0},\\
    \frac{1}{\tau_{\textrm{tr}}(\bn\bn_0)}&=\frac{n_{\text{imp}}\ve^2}{2\pi v_\parallel^3p_0^6}\int\frac{d\Omega}{4\pi}g^\prime_{\pe(\bn-\bn^\prime),\bn_0}
    (1-\cos\theta_{\bn\bn^\prime})\\
    &\equiv
    \frac{1}{\tau}-\frac{1}{\tau_1},\\
        \frac{1}{\tau_{\textrm{tr}_2}(\bn\bn_0)}&=\frac{n_{\text{imp}}\ve^2}{2\pi v_\parallel^3 p_0^6}\int\frac{d\Omega}{4\pi}g^\prime_{\pe(\bn-\bn^\prime),\bn_0}
        (1-\cos^2\theta_{\bn\bn^\prime})\\
        &\equiv
        \frac{1}{\tau}-\frac{1}{\tau_2}.\\
  \end{split}
\end{gather}
The derivation of the scattering rates is sketched in Appendix B.
In accordance to what was said about the validity of the Fermi-liquid approach, we only need the value of the vertex function at $p=p_\ve$.

This allows us to perform a partial momentum integration  in the Dyson equation and write down the equation for the \textit{on-shell} ($p=\pe$) vertex (see  Appendices A and C). It reads
\begin{gather}
  \begin{split}
  \label{dyson-sphere}
  &\mathbf{J}^{ARR}(\bn\bn_0)=\bsigma+\frac{n_{\text{imp}}u_0^2\ve^2}{4\pi v_\parallel^3p_0^6\xi^2}\\
  &\times\int\frac{d\Omega^\prime}{4\pi}\frac{(1+\bn^\prime
  \bsigma)\mathbf{J}^{ARR}(\bn^\prime\bn_0)(1+
  \bn^\prime\bsigma)g^\prime_{\pe(\bn-\bn^\prime),\bn_0}}
  {\frac{2}{\tau(\bn^\prime\bn_0)}-\frac{1}
  {\tau_{\textrm{tr}}(\bn^\prime\bn_0)}}\, ,
  \end{split}
\end{gather}
where we denoted $\mathbf{J}^{ARR} (\ve,\bn\pe)\equiv
\mathbf{J}^{ARR}(\bn\bn_0)$, meaning that the angular dependence
of $\mathbf{J}^{ARR}(\ve,\bn\pe)$ is clearly defined just by the angle between the anisotropy axis and momentum direction (see Fig.~\ref{weyl3} illustrating the scattering angles). The energy argument in the scattering times is suppressed. In the isotropic case, equation~\eqref{dyson-sphere} allows for a simple solution. Since $\mathbf{J}$ is a renormalized current operator, we look for a solution in the form of the polar vector ansatz
\begin{gather}
  \label{ansatz}
  \mathbf{J}^{ARR}(\ve,\bn)=J_1(\ve)\bm{\sigma}+
  J_2(\ve)\bn+J_3(\ve)\bn(\bm{\sigma}\bn).
\end{gather}
 The ansatz~\eqref{ansatz} turns Dyson equation~\eqref{dyson-sphere} into an algebraic one. Changing the integration measure in the expression~\eqref{polar-fin1} for the conductivity $p^2dp=\ve^2 dp/v^2$,
we arrive at the following suitable formula for the conductivity tensor
\begin{gather}
  \label{ansatz-iso}
    \begin{split}
  \sigma_{\alpha\beta}=\frac{e^2}{6\pi v}\delta_{\alpha\beta}\int\limits_{-\infty}^\infty
  \frac{\ve^2d\ve}{2\pi}\de_\ve\tanh\frac{\ve}{2T}
  \frac{J_1(\ve)+J_2(\ve)+J_3(\ve)}{\frac{1}
  {\tau(\ve)}+\frac{1}{\tau_1(\ve)}}\, .
  \end{split}
\end{gather}
The $J_1+J_2+J_3$ term is extracted with the help of  Eqs.~\eqref{dyson-sphere} and \eqref{ansatz}
\begin{gather}
   J_1+J_2+J_3=\frac{\frac{1}{\tau}+\frac{1}{\tau_1}}
   {\frac{1}{\tau}-\frac{1}{\tau_2}}\, .
\end{gather}
The conductivity then takes the typical Fermi-liquid form
\begin{gather}
    \begin{split}
  \sigma_{\alpha\beta}=\frac{e^2}{6\pi v}\delta_{\alpha\beta}\int\limits_{-\infty}^\infty\frac{\ve^2 d\ve}{2\pi}\de_\ve\tanh\frac{\ve}{2T}
 \frac{1}{\frac{1}{\tau(\ve)}-\frac{1}{\tau_2(\ve)}}\, .
  \end{split}
\end{gather}
Now, we write down the temperature dependence of the conductivity in the cases of the long- and short-range disorder potentials.

\subsection{Short-range potential}
In the case $p_0\gg\max\{T/v,\ef/v\}$, the structure factor $g_{\pe(\bn-\bn^\prime)}$ depends only slightly on the scattering angle (between the $\bn$ and $\bn^\prime$ directions) as the scattering is nearly isotropic. This means that $g\big(\pe(\bn-\bn^\prime)/p_0\big)\approx g(0)=1$. The transport scattering time is
\begin{gather}
  \frac{1}{\tau_{\textrm{tr}_2}(\ve)}=\frac{2}{3\tau(\ve)}
  =\frac{n_{\text{imp}}u_0^2}{3\pi v^3 p_0^6}\ve^2\, ,
\end{gather}
and the conductivity reads
\begin{gather}
  \label{cond-iso}
   \sigma_{\alpha\beta}=\sigma\delta_{\alpha\beta},\quad
   \sigma=\frac{1}{2\pi}\frac{e^2v^2p_0^6}{n_{\text{imp}} u_0^2}.
\end{gather}
The conductivity is independent of the chemical potential and temperature in the short-range disorder limit~\cite{SarmaPRB2015}.

\subsection{Long-range potential}
This case corresponds to $p_0\ll\max\{T/v_\parallel,\ef/v_\parallel\}$. Now, the scattering is strongly anisotropic. Then,  $\pe^2(\bn-\bn^\prime)^2=2\pe^2(1-\cos\theta)\approx\pe^2\theta^2$, and
\begin{gather}
  \label{tau-tr2}
  \frac{1}{\tau_{\textrm{tr}_2}(\ve)}=\frac{n_{\text{imp}}u_0^2g_1v}{8\pi}\frac{1}{p_0^2\ve^2}\,,
\end{gather}
where $g_1=\int_0^\infty g(x)x\,dx$  (we assume the convergence of the corresponding integral). The conductivity then reads
\begin{gather}
  \label{cond-long}
  \begin{split}
   \sigma_{\alpha\beta}&=\frac{8}{3}
   \frac{e^2}{n_{\text{imp}}u_0^2g_1}\frac{\ef^4p_0^2}{v^2}
   f\Big(\frac{T}{\ef}\Big)\delta_{\alpha\beta}\,,\\
   f(x)&=\Big[1+2\pi^2x^2+\frac{7\pi^4}{15}x^4\Big]\, .
  \end{split}
\end{gather}
In particular, when the doping level is low, $\ef\ll T$ $(x\gg 1)$,
\begin{gather}
  f(x)\approx\frac{7\pi^4}{15}x^4\, ,
\end{gather}
we have
\begin{gather}\label{T4}
   \sigma_{\alpha\beta}\approx\frac{56}{45}\frac{e^2p_0^2}{n_{\text{imp}} u_0^2 g_1}
  \frac{T^4}{ {v}^2}\delta_{\alpha\beta}\, .
\end{gather}
The temperature dependence $\sigma\sim T^4$ for long-range disorder of arbitrary form can be qualitatively explained in the following manner.
The Born transport scattering rate is $\tau^{-1}_{\text{tr}}\propto (p_0\ve)^{-2}$ and the density of states satisfies $\nu\propto \ve^2$. For $T\gg\ef$, the characteristic energy of the charge carriers becomes $\ve\sim T$. Therefore, the Drude conductivity takes the form
\begin{gather}
\sigma \propto \tau_{\text{tr}}(T)\nu(T)\propto T^4.
\end{gather}

Formula~\eqref{T4} surprisingly gives the same dependence even for Coulomb disorder. As was pointed out in Refs.~\onlinecite{SarmaPRB2015,RodioSyzr_PRB2015}, the screening length for the Coulomb disorder potential is temperature-dependent and $p_0\sim T$. The potential amplitude, however, also depends on temperature $u_0\sim e^2/r\sim e^2 p_0\propto T$ and the $T^4$ dependence survives.

Next, our attention turns to the anisotropic case. As we will see, the
longitudinal and transverse conductivities exhibit different temperature dependence. Also, the uniaxial anisotropy introduces additional geometric factors, which are usually needed by experimentalists. In some cases, we managed to obtain exact results.

\section{Anisotropic case}
\subsection{The Dyson equation and conductivity}
For convenience, we now introduce a modified vertex function  $\mathbf{I}=\mathbf{J}^{ARR}/(2/\tau-1/\tau_{\textrm{tr}})$ and
denote $\mathbf{I}^{ARR} (\ve,\bn\pe)\equiv \mathbf{I}(\bn\bn_0)$
\begin{gather}
  \begin{split}
  \label{dyson-sphere1}
  &\mathbf{I}(\bn\bn_0)\left(\frac{2}{\tau(\bn\bn_0)}-
  \frac{1}{\tau_{\textrm{tr}}(\bn\bn_0)}\right)=
  \bsigma+\frac{n_{\text{imp}}u_0^2\ve^2}{4v_\parallel^3p_0^6\pi\xi^2}\\
  &\times\int\frac{d\Omega^\prime}{4\pi}(1+\bn^\prime\bsigma)
  \mathbf{I}(\bn^\prime\bn_0)(1+\bn^\prime\bsigma)
  |g^\prime_{\pe(\bn-\bn^\prime),\bp_0}|^2\, .
  \end{split}
\end{gather}

The solution to the Dyson equation is sought in the form of the most general  polar vector composed of $\bsigma,\ \bn$ and $\bn_0$:
\begin{gather}
  \label{ans1}
  \begin{split}
  \mathbf{I}=I_1\bsigma+I_2\bn+I_3\bn(\bn\bsigma)+
  I_4\bn_{0}(\bn\bsigma)\\
  +I_5\bn(\bn_0\bsigma)+I_6\bn_0+I_7\bn_0(\bn_0\bsigma)\, .
  \end{split}
\end{gather}

\begin{figure}[t]
\includegraphics[width=0.8\columnwidth]{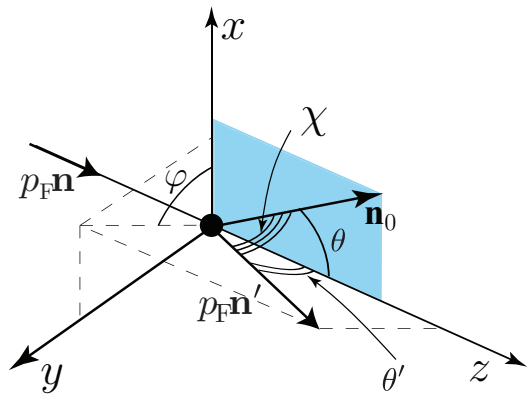}
 \caption{\label{weyl3}(Color online)
Scattering event: $\bq$ is the momentum of the incoming particle,
$\bn_0$ is the axis of the uniaxial anisotropy. For simplicity, the
coordinate system is chosen such that $\bq,\ \bn_0$ span the $xz$
plane; $\bp$ is the scattered momentum, $\theta^\prime$ and $\vf$
are the polar and azimuthal scattering angles, while $\chi$ is the
scattering angle with respect to the anisotropy axis.
          }
\end{figure}

Plugging~\eqref{ans1} into expression~\eqref{polar-fin} for the conductivity tensor, we obtain
\begin{gather}
   \label{cond-fin}
  \begin{split}
    \sigma_{\alpha\beta}(\ve)=\frac{e^2\ve^2}{2\pi^2v_\parallel}
   \int\frac{d\Omega}{4\pi}\int\Big[\Phi(\bn\bn_0)n_\alpha n_\beta\\
   +
   \frac{1}{2}\Psi(\bn\bn_0)(n_{0\alpha}n_\beta+
   n_{\alpha}n_{0\beta})\Big]\, ,
  \end{split}
\end{gather}
where $\Phi=I_1+I_2+I_3+I_5 x,\ \Psi=I_4+I_6+I_7 x,$ and $x = \cos\theta$.

\subsection{Short-range potential}
\subsubsection{$\delta$-correlated potential}

A weak momentum dependence of the potential leaves it essentially isotropic, because it does not depend on the momentum. Therefore, the answer for the $\delta$-correlated potential (its characteristic momentum is $p_0=\infty$) is obtained immediately
\begin{gather}
 \sigma_\parallel=\sigma,\ \ \sigma_\bot=\xi^2\sigma\, ,
\end{gather}
where the conductivity $\sigma$ is defined in~\eqref{cond-iso} with $v=v_\parallel$.

\subsubsection{Finite-range potential corrections}
Now it is clear that the anisotropy in the rescaled basis enters the conductivity expression only as a $\max\{\ef,T\}/p_0$ correction.
The first correction in the $\max\{\ef,T\}/p_0$ series is easy to compute, expanding the Dyson equation. The spectrum also acquires anisotropic corrections. They are proportional to the disorder potential amplitude and are irrelevant for sufficiently weak potential.
We expand the disorder form factor~\eqref{form-factor} according to
\begin{gather}
  \label{form1}
  g(x)\approx 1+\de g(0) x,\ \ x=\frac{p^2}{p_0^2}\ll 1.
\end{gather}
Using Eq.~\eqref{cond-fin} and solving the Dyson equation~\eqref{dyson-sphere1} with the ansatz~\eqref{ans1}, we obtain the corresponding correction to the conductivity (the details of the computation are summarized in Appendix E)
\begin{gather}
 \label{cond-short0}
  \begin{split}
             \sigma_\parallel=\sigma\Big[1-\de g(0)
             \frac{1+4\xi^{-2}}{5}\frac{\pi^2T^2/3+\ef^2}{(v_\parallel p_0)^2}\Big]\,,\\
              \sigma_\bot=\sigma\xi^2\Big[1-\de g(0)
              \frac{7+8\xi^{-2}}{15}\frac{\pi^2T^2/3+\ef^2}{(v_\parallel p_0)^2}\Big]\,.
  \end{split}
\end{gather}
It is worthwhile to note that the anisotropy manifests itself in the temperature-dependent conductivity ratio
 \begin{gather}   \label{result}
     \frac{\sigma_\parallel\xi^2}{\sigma_\bot}=
     1-\frac{4}{15}(1+\xi^{-2})\de g(0)\frac{\pi^2 T^2/3+\ef^2}{(v_\parallel p_0)^2}\, .
  \end{gather}
Equation~\eqref{result} is one of the central results of the paper. We have just found that the conductivity components in the parallel and transverse directions exhibit different temperature behavior. Although Eq.~\eqref{result} is obtained in the limit $p_0\ll\max\{T/v,\ef/v\}$, we guess that the temperature dependence of the $\sigma_\parallel\xi^2/\sigma_\bot$ ratio should become more pronounced when $p_0\sim \max\{T/v,\ef/v\}$.

\subsection{Long-range potential}
\begin{figure}[t]
\includegraphics[width=0.6\columnwidth]{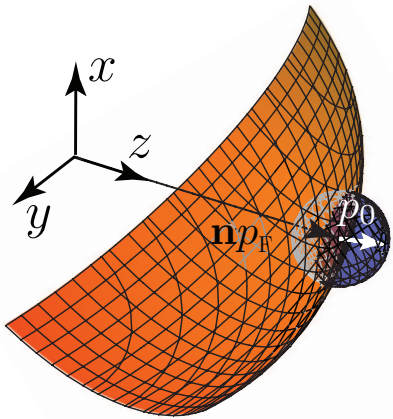}
 \caption{\label{weyl3b}(Color online)
A part of the Fermi sphere (shown in orange). The small dark sphere represents the possible change of momentum due to scattering.
          }
 \end{figure}

The long-range potential $u(\bp)$ corresponds to the case
 $p_0\ll \max\{\ef/v_\parallel,T/v_\parallel\}$.
Now the ratio
\begin{gather}
   \kappa_\ve=\frac{p_0}{\pe}\ll1
\end{gather}
becomes an additional small parameter of the problem.
The disorder potential does not change much the momentum of an incoming  particle $\delta p\sim p_0\ll \pe$. Therefore, hereafter we can consider small scattering angles wherever necessary (see Fig.~\ref{weyl3b}).

The calculations are rather cumbersome. They are presented in Appendix F. The Dyson equation is turned into two coupled differential equations
\begin{gather}
 \begin{split}
   \label{dyson-fin1}
   &\Phi^{\prime\prime}(1-x^2)+x\Phi^\prime(-3-\xi^2+x^2(\xi^2-1))\\
   -&\Phi(1+\xi^2-x^2(\xi^2-1))=-\frac{1}{2}(\xi^2-(\xi^2-1)x^2)^{3/2}, \\
   &\Psi^{\prime\prime}(1-x^2)-x[(\xi^2+1)-x^2(\xi^2-1)]
   \Psi^\prime=-2\Phi^\prime\,,
  \end{split}
\end{gather}
where, for brevity, we switched to dimensionless functions
$\frac{\kappa_\ve^2\xi^2}{4}\frac{g_1}{g_0\tau_0}(\Phi,\Psi)\rightarrow(\Phi,\Psi)$ and $g_0=\int_0^\infty g(x)\,dx$. Naturally, the problem of boundary conditions immediately comes on stage. As will be shown below, it is possible to circumvent it in a number of important limiting cases. The Dyson equation for $\Psi$ determines the $\Psi(x)$ function up to an arbitrary constant. This, however, does not cause any difficulty for the conductivity, because $\Psi(\bn\bn_0)$ enters the integral~\eqref{cond-fin} with $\bn$ as a multiplier and forms an odd function of the polar angle.

The resultant expressions for
$\sigma_\parallel$ and $\sigma_\bot$ read
\begin{gather}
  \begin{split}
  \sigma_\parallel(T)&=\sigma_1f\Big(\frac{T}{\ef}\Big)\int\limits_{-1}^1
  [\Phi(x)x^2+\Psi(x)x]\,dx,\\
  \sigma_\bot(T)&=\frac{\xi^2\sigma_1}{2}f\Big(\frac{T}{\ef}\Big)
  \int\limits_{-1}^1\Phi(x)(1-x^2)\,dx,\\
   \sigma_1&=\frac{8e^2 \ef^4p_0^2}{n_{\text{imp}} g_1 u_0^2v_\parallel^2}\, ,
  \end{split}
 \end{gather}
where the function $f(x)$ is defined in~\eqref{cond-long}.

Next, we explore the geometrical implications of anisotropy in the three limiting cases.

\subsection{Limiting cases of anisotropy}

\subsubsection{Easy plane, $\xi\gg1$}

The first one is the case of strong anisotropy, when $\xi\gg1$
($v_\bot\gg v_\parallel$). Then, the Dyson equations~\eqref{dyson-fin1} reduce to the first-order differential equation
\begin{gather}
  \begin{split}
    (\Phi x)^\prime=\frac{\xi}{2}\sqrt{1-x^2}\, ,\\
       \Psi\sim
       \hbox{const}+\mathcal{O}\Big(\frac{1}{\xi^2}\Big)\, .
   \end{split}
 \end{gather}
The constant in $\Psi(x)$ is irrelevant as was mentioned earlier;
while the solution, which is analytic in the interval $x\in[-1,1]$, reads
\begin{gather}
   \Phi(x)=\frac{\xi}{4}\Big(\sqrt{1-x^2}+\frac{\arcsin x}{x}\Big).
 \end{gather}
 The conductivity can be then written as
  \begin{gather}
   \label{long-res}
   \begin{split}
   \sigma_\parallel(T)&=
   \frac{3\pi\xi}{32}\sigma_1f\Big(\frac{T}{\ef}\Big),\\
   \sigma_\bot(T)&=\frac{\pi\xi^3}{32}\sigma_1f\Big(\frac{T}{\ef}\Big)
   \Big(\frac{1}{2}+4\ln2\Big).
   \end{split}
\end{gather}

The conductivity ratio becomes
\begin{gather}
   \frac{\sigma_\parallel\xi^2}{\sigma_\bot}=\frac{3}{\frac{1}{2}+4\ln 2}+\mathcal{O}(\xi^{-1}),\ \ \xi\rightarrow\infty.
\end{gather}

\subsubsection{Weak anisotropy, $\xi\approx 1$}

Now we turn to the case of $v_\bot\approx v_\parallel$. Let
$\xi=1+\delta\xi,\ \ \delta\xi\ll1$. We expand the vertex function
according to $\Phi=\frac{1}{4}+\delta\Phi(x)$ (the value $1/4$
corresponds to the isotropic scattering). The corresponding equation
reads
 \begin{gather}
   \delta\Psi^{\prime\prime}(1-x^2)-
   4x\delta\Phi^\prime-2\delta\Phi=-\delta\xi(1-x^2)\, .
\end{gather}
The solution analytic in $x\in[-1,1]$ has the form

 \begin{gather}
  \begin{split}
   \Phi(x)=\frac{1}{4}-\frac{\delta\xi}{12}(x^2-5),\\
   \Psi(x)=-(\delta\xi)\frac{x}{6}\,.
  \end{split}
\end{gather}
Then, the conductivity reads
\begin{gather}
  \begin{split}
   \sigma_\parallel(T)=\frac{1}{6}\sigma_1f\Big(\frac{T}{\ef}\Big)\Big(1+
   \frac{4}{15}\delta\xi\Big),\\
   \sigma_\bot(T)=\frac{1}{6}\sigma_1 f\Big(\frac{T}{\ef}\Big)\Big(1+
   \frac{34}{15}\delta\xi\Big).\ \
  \end{split}
\end{gather}
It is worth noting that to the first order in the deviation from anisotropy,
the ratio
\begin{gather}
   \frac{\sigma_\parallel\xi^2}{\sigma_\bot}=1+
   \mathcal{O}(\delta\xi^{2}),\ \ \delta\xi\ll1.
\end{gather}
does not exhibit any shift.

\subsubsection{Easy axis, $\xi\ll 1$}

In this case, we are unable to obtain exact coefficients. However, it is enough to say that at $\xi=0$ there exist finite solutions $\Phi(x)$ and $\Psi(x)$, which lead to some finite integrals with $x^2$ and $x$ in the $[-1,1]$ range. Then, we have a qualitative answer
\begin{gather}
 \begin{split}
   \sigma_\parallel&=\sigma_1 f\Big(\frac{T}{\ef}\Big)c_1,\\
   \sigma_\bot&=\sigma_1\xi^2f\Big(\frac{T}{\ef}\Big)c_2\, ,
  \end{split}
\end{gather}
where $c_1$ and $c_2$ are constants of the order of unity. Then, the conductivity ratio takes the form
\begin{gather}
   \frac{\sigma_\parallel\xi^2}{\sigma_\bot}=
   \frac{c_1}{c_2}+\mathcal{O}(\xi^{2}),\ \ \xi\ll1.
\end{gather}

\begin{figure}[t]
\includegraphics[width=0.95\columnwidth]{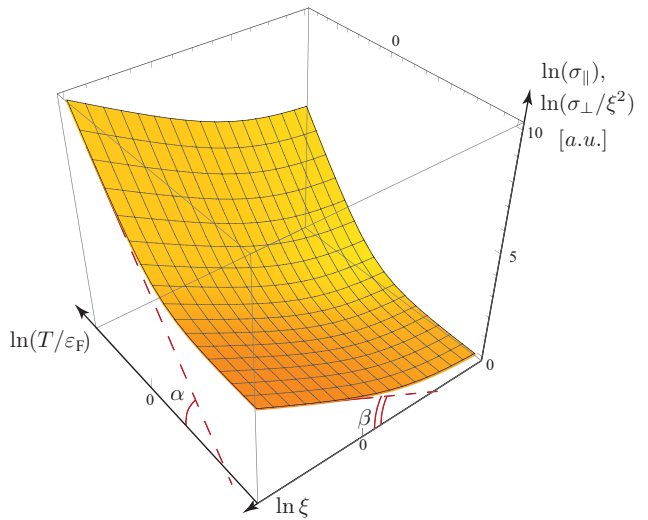}
 \caption{\label{depend1}(Color online) Conductivities $\sigma_{\parallel,\bot}(T,\xi)$ as a function of the anisotropy parameter $\xi$; $\tan\alpha=4$,\ \ $\tan\beta=1$.
          }
 \end{figure}

The above results are summarized in Fig.~\ref{depend1}.

\section{Conclusions}

The results of this work can be summarized as follows. We have rigorously studied the conductivity of isotropic and anisotropic weakly disordered WSM in two important limits: short- and long-range disorder ($p_0\ll$ or $\gg \max\{T,\ef\}/v_\parallel$). With the help of the diagrammatic approach, we have been able to justify the applicability of the Fermi-liquid theory to WSM. The disorder potential was assumed to have a general form.

We have found that short- and long-range disorder leads to different temperature dependences of the conductivity. In the case of short-range disorder, we have discovered that uniaxial anisotropy leads to even different temperature dependences of the longitudinal and transverse conductivities.
In contrast, the long-range disorder yields identical temperature dependences of conductivity components, $\sigma_{\parallel,\bot}\sim T^4$, for $T\gg\ef$.

We have  also explored the dependence of the conductivity tensor on the anisotropy parameter $\xi$ and established general scaling relations in the cases of strong and weak anisotropy. We have managed to compute analytically the geometric factors for the conductivity tensor in the limit of strong $\xi\gg$ and weak $\xi\approx 1$ anisotropy.

The recent experimental data on WSM makes it possible to estimate the
Fermi velocities and Fermi energies for typical samples. For example, for both Na$_3$Bi~\cite{Liu_Science2014} and Cd$_3$As$_2$~\cite{Liu_NatMat2014}, the Fermi velocity ratio is $\xi \approx 4$; thus the regime $\xi\gg 1$ is realized and the anisotropy of the conductivity should be clearly pronounced ($\sim\xi^2$). The Fermi energy takes values in the 100--1000 K range. Therefore, depending on the disorder correlation length, results~\eqref{long-res} or \eqref{result} should be applicable.

The developed approach and the obtained results provide a good basis for further progress in the field of transport phenomena in 3D systems with Dirac points in their energy spectrum.

\section*{Acknowledgments}

We are grateful to S.V. Syzranov for a critical reading of the manuscript and numerous useful comments, as well as to I.S. Burmistrov, A.S. Ioselevich, V.A. Kagalovsky, and A.G. Semenov for helpful discussions. This work was supported by the Russian Foundation for Basic Research (projects 14-02-00276, 14-02-00058, 15-02-02128, and 12-02-92100-JSPS), the Russian Science Support Foundation, the RIKEN iTHES Project, the MURI Center for Dynamic Magneto-Optics, and a Grant-in-Aid for Scientific Research (A).

\appendix

\section{Expression for the polarization operator
\label{Sec:AppA}}

In order to work out the polarization-operator diagram in
Fig.~\ref{diag0} (left or right), we follow  the scheme proposed by
Eliashberg~\cite{eliashberg}. To build an analytical continuation of the expression for the Matsubara polarization operator, we need to establish the analytical properties of the vertex function  $J_\alpha(z,z+i\omega_n,i\omega_n;\bp)$ in the whole domain of the
complex variable $z$. This is done via the Lehman representation and it was discussed in detail in, e.g., Ref.~\onlinecite{rod-bur-ios}.

The conclusion is that the domain of analyticity of $J_\alpha(z,z+i\omega_n,i\omega_n;\bp)$ is a complex plane with two horizontal cuts:\ $\hbox{Im}(z+i\omega)=0$\ and\
$\hbox{Im}(z)=0$. Since we need a {\it retarded} vertex function,
we put\ $\omega_n>0$. Next, the three vertex functions are defined
in accordance with the structure of the cuts
 \begin{gather}
  \label{gamma-define}
   \begin{split}
       &J_\alpha^{RRR}(z,z+i\omega,i\omega)\quad
       {\rm if}\quad\hbox{Im} z>0,\\
       &J_\alpha^{ARR}(z,z+i\omega,i\omega)\quad
       {\rm if}\quad-i\omega_n<\hbox{Im} z<0,\\
       &J_\alpha^{AAR}(z,z+i\omega,i\omega)\quad
       {\rm if}\quad\hbox{Im}
       z<-i\omega_n.
   \end{split}
\end{gather}
The general expression for\
$\Pi_{\alpha\beta}(i\omega_n)$\ then becomes
\begin{figure}[t]
  \includegraphics[width=60mm]{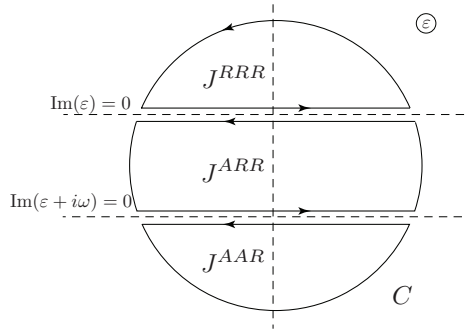}
   \caption{ Contour for polarization operator\ $\Pi(\omega)$.
   \label{figure11}
          }
 \end{figure}
\begin{widetext}
\begin{gather}
 \begin{split}
   &\Pi_{\alpha\beta}(i\omega_n)=-T\sum_{\varepsilon_k}
   J_{\alpha}(i\varepsilon_k,i\varepsilon_k+i
   \omega_n,i\omega_n;\bp)G(i\varepsilon_k+i\omega_n,\bp)j_\beta
   G(i\varepsilon_k,\bp)\\
   &=-\oint_C
   \frac{d\varepsilon}{4\pi i}\tanh\frac{\varepsilon}{2T}
   J_\alpha(\varepsilon,\varepsilon+i\omega_n,i\omega_n;\bp)
   G(\varepsilon+i\omega_n,\bp)j_\beta G(\varepsilon,\bp).
 \end{split}
\end{gather}
The contour\ $C$\ is shown in Fig.~\ref{figure11}. As usual, the
integral over the large circle vanishes and we are left with integrals
over different branches
\begin{gather}
 \begin{split}
   &\Pi_{\alpha\beta}(i\omega_n)=\int^{\infty}_{-\infty}
   \frac{d\varepsilon}{4\pi i}\tanh\frac{\varepsilon}{2T}
  \Big\{J_\alpha^{RRR}(\varepsilon,\varepsilon+i\omega_n,i
  \omega_n;\bp)G^R(\varepsilon+i\omega_n,\bp)j_\beta G^R(\varepsilon,\bp)-\\
   &-J_\alpha^{ARR}(\varepsilon,\varepsilon+i\omega_n,i\omega_n;
   \bp)G^R(\varepsilon+i\omega_n,\bp)j_\beta G^A(\varepsilon,\bp)+
  J_\alpha^{ARR}(\varepsilon-i\omega,\varepsilon,i\omega_n;
  \bp)G^R(\varepsilon,\bp)j_\beta G^A(\varepsilon-i\omega_n.\bp)-\\
   -&J_\alpha^{AAR}(\varepsilon-i\omega_n,\varepsilon,i\omega_n;\bp)
   G^A(\varepsilon,\bp)j_\beta G^A(\varepsilon-i\omega_n,\bp)\Big\}.
 \end{split}
\end{gather}
Making the analytic continuation\ $i\omega_n\rightarrow\omega+i0$, we obtain
\begin{gather}
    \label{polarization0}
   \Pi^{R}_{\alpha\beta}(\omega)= -v_\parallel^2\tr\int\frac{d\varepsilon}
   {4\pi i}\frac{d\bp}{(2\pi)^3}
   \Big\{G^{A}(\varepsilon,\bp) J_{\alpha}^{ARR}(\varepsilon,
   \varepsilon+\omega,\omega;\bp)G^{R}(\varepsilon+\omega,\bp) \sigma_\beta
   \Big[\tanh\frac{\varepsilon+\omega}{2T}-\tanh\frac{\varepsilon}{2T}\Big] \\
+
G^R(\varepsilon,\bp)J_{\alpha}^{RRR}(\varepsilon,\varepsilon+\omega,
\omega;\bp)G^R(\varepsilon+\omega;\bp)
\sigma_\beta
    \tanh\frac{\varepsilon}{2T}
   -G^A(\varepsilon,\bp) J_{\alpha}^{AAR}(\varepsilon,\varepsilon+
   \omega,\omega;\bp)G^A(\varepsilon+\omega,\bp) \sigma_\beta
   \tanh\frac{\varepsilon+\omega}{2T} \notag
   \Big\}.
 \end{gather}
Here, the vertex functions $J_\alpha$ are defined diagrammatically in Fig.~\ref{diag0}c. They obey the Dyson equation presented in a
diagrammatic form in Fig.~\ref{diag2}. The disorder is static and
does not cause a change of frequency in the diagrammatic loops.
Thus, the Dyson equation takes an especially elegant form
 \begin{gather}
  \label{vertex0}
       J_\alpha(i\varepsilon_k,i\varepsilon_k+i\omega_n,i
       \omega_n;\bq)=\sigma_\alpha+\frac{n_{\text{imp}}u_0^2}
       {p_0^6}\int\frac{d\bp}{(2\pi)^3}
       g^\prime_{\bq-\bp,\bn_0} G(i\varepsilon_k,\bp)
       J_\alpha(i\varepsilon_k,i\varepsilon_k+i\omega_n,i
       \omega_n;\bq) G(i\varepsilon_k+i\omega_n,\bp)\,.
\end{gather}
We are interested in the zero-frequency response of the system
and set the external frequency $\omega=0$. After the analytic continuation, the
vertex function is split into  $RRR,\ AAR$ and $ARR$ parts
\begin{gather}
  \label{vertex1a}
  \begin{split}
       &J^{ARR}_\alpha(\ve,\bq)=\sigma_\alpha+\frac{n_{\text{imp}}u_0^2}{p_0^6\xi^2}
       \int\frac{d\bp}{(2\pi)^3}g^\prime_{\bq-\bp,\bn_0}
        G^A(\varepsilon,\bp) J^{ARR}_{\alpha}
       (\varepsilon,\bp) G^R(\varepsilon,\bp)\,,\\
       &J^{RRR}_\alpha(\ve,\bq)=\sigma_\alpha+\frac{n_{\text{imp}}u_0^2}{p_0^6\xi^2}
       \int\frac{d\bp}{(2\pi)^3}g^\prime_{\bq-\bp,
       \bn_0} G^R(\varepsilon,\bp) J^{RRR}_{\alpha}
       (\varepsilon,\bp) G^R(\varepsilon,\bp)\,,\\
       &J^{AAR}_\alpha(\ve,\bq)=\sigma_\alpha+\frac{n_{\text{imp}}u_0^2}{p_0^6\xi^2}
       +\int\frac{d\bp}{(2\pi)^3}g^\prime_{\bq-\bp,\bn_0}
   G^A(\varepsilon,\bp) J^{AAR}_{\alpha}
       (\varepsilon,\bp) G^A(\varepsilon,\bp)\,. \\
  \end{split}
\end{gather}
\end{widetext}
The learned reader already knows that only $J^{ARR}$ (so called
singular) vertex undergoes a strong (non-perturbative)
renormalization due to the interaction with impurities, while the
other two exhibit weak perturbative corrections. The reason is, of
course, the position of the poles of the Green's functions. Since we
claim some rigor, we discuss this issue in detail in Appendix
B.

\section{One-loop structure of the Green's function}
\subsection{General expression}
The general expression for the one-loop self-energy reads
 \begin{gather}
  \label{im-self1}
  \begin{split}
  \Sigma^R(\ve,\bq)&=\frac{n_{\text{imp}}u_0^2}{p_0^6\xi^2}
  \int\frac{d\bp}{(2\pi)^3}\frac{\ve+v_\parallel\bp\bsigma}
  {(\ve+i0)^2-v_\parallel^2\bp^2}g^\prime_{\bq-\bp,\bn_0}
  \end{split}
 \end{gather}
 and its imaginary part
 \begin{gather}
  \label{im-self2}
  \begin{split}
  \Im\Sigma^R(\ve,\bq)&=-\frac{n_{\text{imp}}u_0^2\ve^2}{4\pi v^3_\parallel p_0^6\xi^2}\int(1+\bn\bsigma)
 g^\prime_{\bq-\frac{\ve\bn}{v_\parallel},\bn_0}\frac{d\Omega}{4\pi}\,.
  \end{split}
\end{gather}
The self energy has the following tensor form
\begin{gather}
   \Sigma^R(\ve,\bq)=\alpha(\ve,\bq)+\bm{\beta}(\ve,\bq)
   \bsigma-\frac{i}{2\tau(\ve,\bq)}-\frac{i\bsigma\mathbf{s}
   (\ve,\bq)}{2\tau_1(\ve,\bq)}\, ,
\end{gather}
where $\bm{s}$ and $\bm{\beta}$ are vectors in the ($\bq,\bn_0$) plane and
\begin{gather}
  \frac{\alpha}{\ve},\frac{|\bm{\beta}|}
  {\ve}\sim\Big(\frac{p_0v_\parallel}{\ve}
  \Big)^2\frac{\hbox{min}\{p_0v_\parallel,\ve\}}
  {\ve}\frac{1}{\ve\tau}\ll1\, .
\end{gather}
For the case of isotropic potential, the anisotropic short range potential $(p_0\gg\pe)$ and
anisotropic long range potential $(p_0\ll\pe)$, vector $\bm{s}\equiv\bn$ and the scattering rates $\tau$ and $\tau_1$ are immediately extracted
and presented in~\eqref{sc}.

As is seen from the structure of the integrals~\eqref{im-self1}
and \eqref{im-self2}
 \begin{gather}
  \label{parity}
  \begin{split}
  \tau(\ve,-q,\theta)&=\tau(\ve,q,\theta);\\
    \frac{\bm{s}(\ve,-q,\theta)}{\tau_1(\ve,-q,
    \theta)}&=-\frac{\bm{s}(\ve,q,\theta)}{\tau_1(\ve,q,\theta)}.
  \end{split}
\end{gather} Here $\theta$ is the angle between $\bq$ and $\bn_0$
and the parity relation~\eqref{parity} follows from the measure
invariance under the change $\theta\rightarrow\pi-\theta,\
\vf\rightarrow\vf+\pi$.

Now we may write the Green's function in the standard way
\begin{gather}
  \label{green-exact}
   G^R(\ve,\bq)=\frac{\ve-\alpha+\frac{i}{2\tau}+
   \bsigma(\bq+\bm{\beta}-\frac{i\mathbf{s}}{2\tau_1})}
   {(\ve-\alpha+\frac{i}{2\tau})-(\bq+\bm{\beta}-
   \frac{i\mathbf{s}}{2\tau_1})^2}\, .
\end{gather}

In a typical computation, we need to integrate products of the type
$G^R(\ve,q)G^R(\ve,q),\ G^A(\ve,q)G^A(\ve,q)$ and
$G^A(\ve,q)G^R(\ve,q)$ over the momentum $q\in(-\infty,\infty)$.
Therefore, we need to know the position of the poles of the Green's
function in the $q$-domain.

\begin{figure}[t]
  \includegraphics[width=85mm]{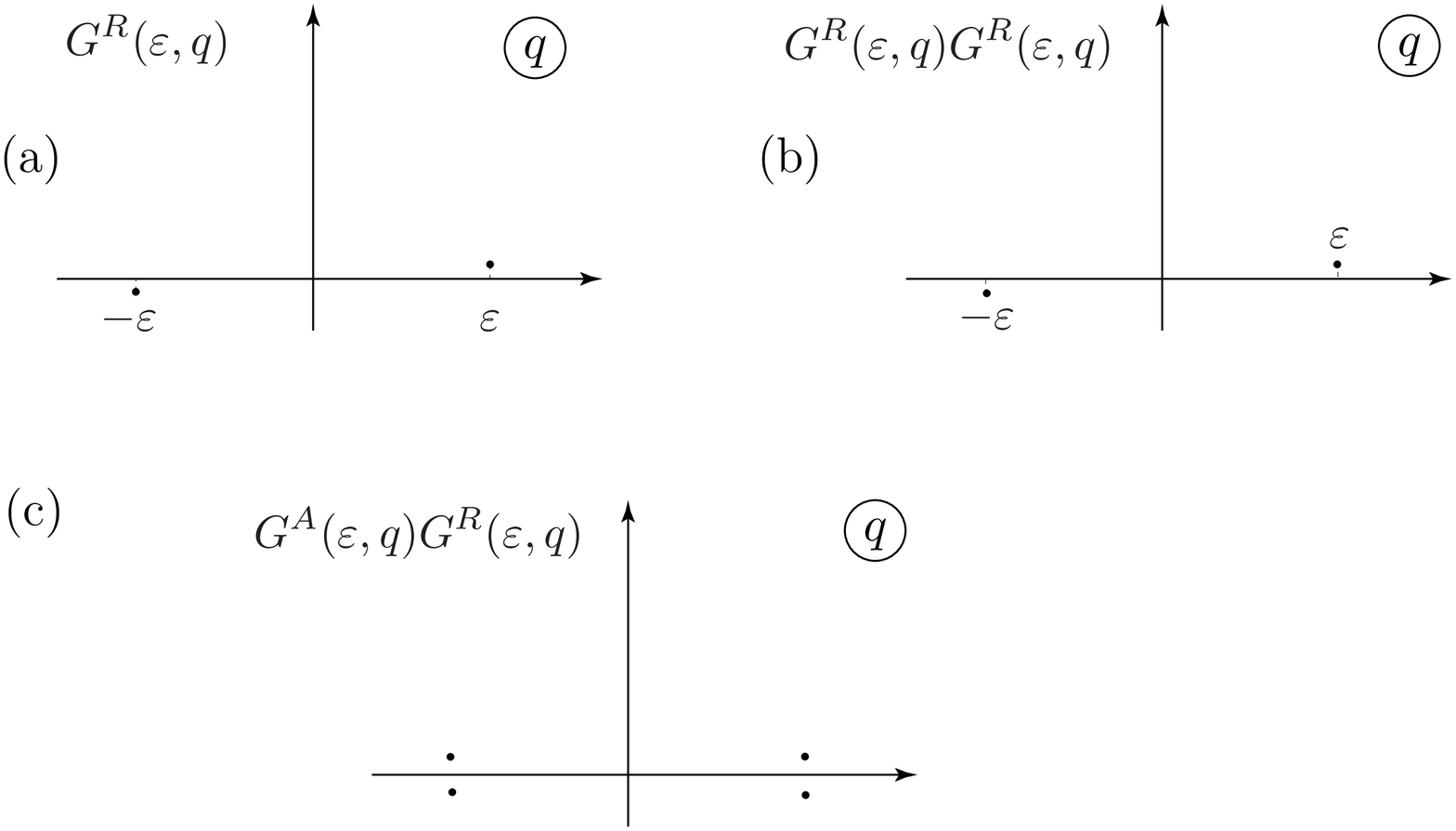}
   \caption{Poles of (a) $G^R(\ve,q)$,\ (b) $G^R(\ve,q)G^R(\ve,q)$, and (c) $G^A(\ve,q)G^R(\ve,q)$, in the $q$ plane.
   \label{poles}
          }
\end{figure}

The retarded function $G^R(\ve,\bq)$ obeys the causality condition,
which means that its poles, defined by the equation
 \begin{gather}
  \label{spec1}
  \hbox{det}[\ve-v_\parallel\bq\bsigma-\Sigma^R(\ve,\bq)]=0\, ,
 \end{gather}
lie in the lower half-plane of the complex variable $\ve$, i.e.

\begin{gather}
  \label{analyt}
  \frac{1}{\tau(\ve,\ve,\theta)}+\frac{\bn\bm{s}(\ve,\ve,\theta)}
  {\tau_1(\ve,\ve,\theta)}>0
\end{gather}
for any $\ve$ and $\theta$; $\bn$ stands for $\bq/q$.
Since we need the position of the poles in the $q$-plane, we
solve the spectral equation~\eqref{spec1} for $q$ and obtain two
roots. Their imaginary parts are
 \begin{gather}
   \Im q(\ve)=\pm\Big[\frac{1}{2\tau(\ve,\pm\ve,\theta)}\Big]
  -\frac{\bm{s}(\ve,\pm\ve,\theta)\bn}{2\tau_1(\ve,\pm\ve,\theta)}\,.
 \end{gather}
Using~\eqref{parity} and~\eqref{analyt}, we see  that the position
of the poles of $G^R(\ve,\bq)$ in the complex plane $q$ is always such as depicted in Fig.~\ref{poles}a. The poles of $G^R(\ve,q)G^R(\ve,q)$ and  $G^A(\ve,q)G^R(\ve,q)$ are sketched in Figs.~\ref{poles}b and \ref{poles}c. Now, we see that while integrating the product $G^RG^R$ or $G^A G^A$ over $q$ the contour of integration can be deformed to pass far from the poles. When integrating the product $G^AG^R$, the contour is squeezed between the nearby poles and the deformation is no longer possible, leading to their enhanced contribution.

\section{Dyson equation}
\subsection{Structure of $\mathbf{J}^{ARR}(\ve,p)$\label{App:Asympt}}

In order to find the conductivity, we need to solve the Dyson equation for the vertex function $J^{ARR}(\ve,p)$ at $p=\ve/v_\parallel$ as well as explore its asymptotic behavior at $p\rightarrow\infty$ (to justify the Fermi-liquid approach). The asymptotic behavior is easier to extract focusing now on the isotropic potential. This allows us to capture all the principal details and avoid unnecessary complications due to uniaxial anisotropy.

\begin{figure}[t]
\centering
  \includegraphics[width=0.6\columnwidth]{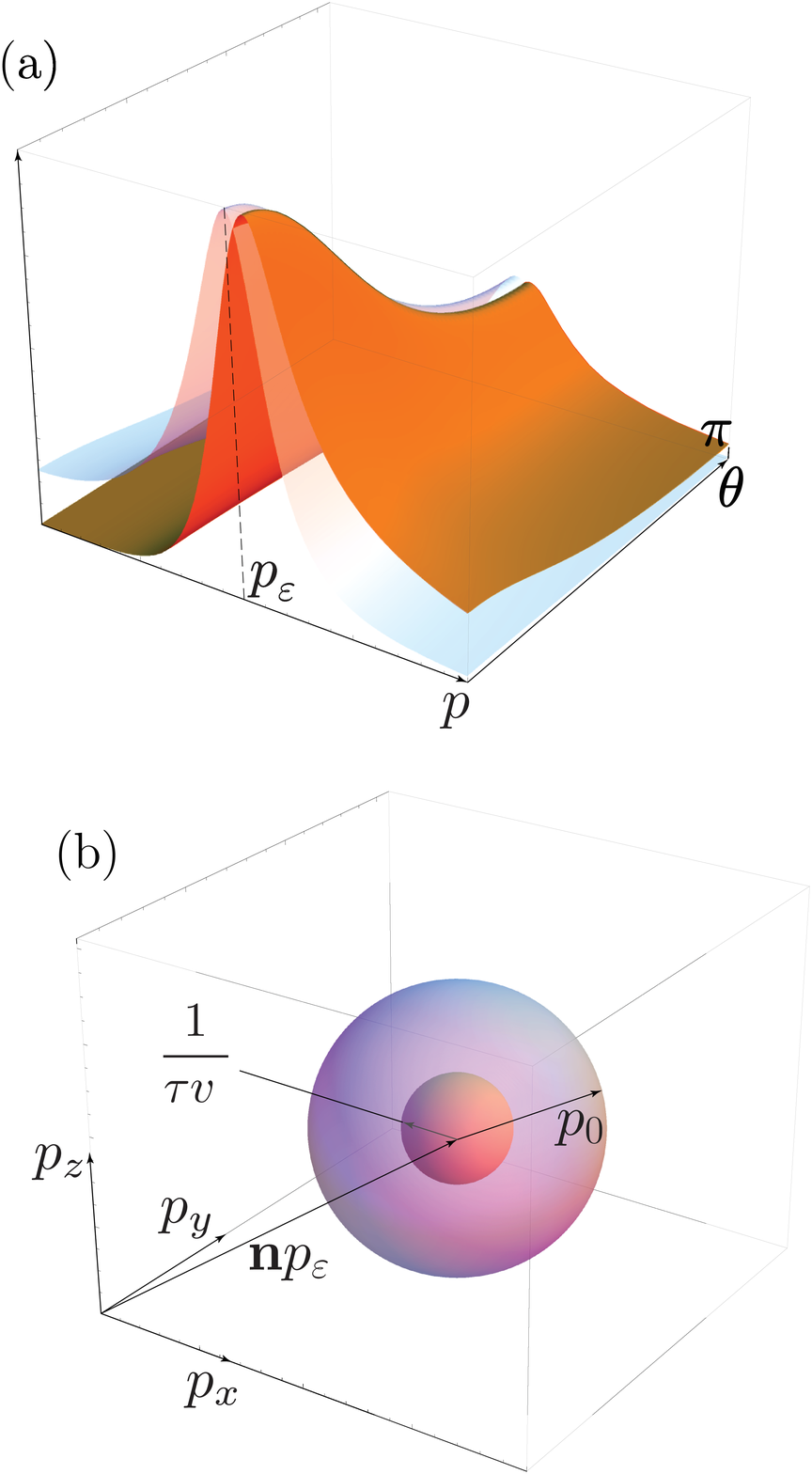}
   \caption{\label{conv1}(Color online)
(a) Approximate integrand $G^R(p,\theta)G^A(p)\pe^2g_{\pe(\bn-\bn^\prime)}$
(transparent) versus the exact one $G^R(p,\theta)G^A(p,\theta)g_{\bq-\bp}p^2$(solid), for a typical potential. (b) The range of integration over $p$. The transparent region (radius $p_0$) corresponds to the smaller contribution.
          }
\end{figure}

The ansatz for the vertex comprises  only three vertices (Eq.~\ref{ansatz}). 
However, in formula~\eqref{ansatz} the vertex is taken on a mass shell at $\bp=\pe=\ve/v$.
Here we consider a complete vertex
\begin{gather}
  \mathbf{J}(\ve,\bq) =J_1(\ve,q)\bsigma+J_2(\ve,q)\frac{\pe\bq}{q^2}+J_3(\ve,q)\frac{\bq(\bq\bsigma)}{q^2}
\end{gather}
For brevity, we will omit the energy symbol $\ve$ in $J(\ve,q)\equiv J(q)$ to restore it later. Using the identity $[\ve+v\bp\bsigma]\bsigma[\ve+v\bp\bsigma]=(\ve^2-v^2p^2) \bsigma+2v\bp[\ve+v\bp\bsigma]$, we obtain the equation
\begin{widetext}
 \begin{gather}
  \label{vertex-calc}
  \begin{split}
  &J_1(q)\bsigma+J_2(q)\frac{\pe\bq}{q^2}+J_3(q)\frac{\bq(\bq\bsigma)}{q^2}=
  \bsigma+ \frac{n_{\text{imp}}u_0^2}{p_0^6}\\
  &\times
     \int\frac{d\bp}{(2\pi)^3}g_{\bq-\bp}
       \frac{J_1(p)\big[(\pe^2\hskip-1mm-\hskip-0.5mm p^2)
       \bsigma+2\bp(\pe+\bp\bsigma)]+J_2(p)\frac{\pe\bp}
       {p^2}\big[\pe^2\hskip-0.5mm+\hskip-0.5mm p^2+2\pe(\bp\bsigma)\big]+
       J_3(p)\frac{\bp}{p^2}
  \Big[(\pe^2\hskip-0.5mm+\hskip-0.5mm p^2)(\bp\bsigma)
  \hskip-0.5mm+\hskip-0.5mm 2\pe p^2\Big]}{\Big[\Big(\ve+\frac{i}
  {\tau(\ve,p)}\Big)^2-\Big(vp-\frac{i}{\tau_1(\ve,p)}\Big)^2\Big]
       \Big[\Big(\ve-\frac{i}{\tau(\ve,p)}\Big)^2-\Big(vp+\frac{i}
       {\tau_1(\ve,p)}\Big)^2\Big]} .\\
  \end{split}
 \end{gather}
\end{widetext}
For the integral~\eqref{vertex-calc} to converge, the vertices
$J_{1,2,3}(p)$ must be bounded when $p\rightarrow\infty$. As in expression for the conductivity, the integrand has a sharp peak at $p=\pe$ of width $\Delta p\sim 1/(v\tau)$ near the Fermi surface. Generally, we cannot restrict the computation of the integral by expanding the integral near the Fermi surface. Such an expansion leads to a significantly deformed integrand (see Fig.~\ref{conv1}(a)). First, however we discuss the vertex at $p=\pe$.

\subsection{On-shell vertex: $J^{ARR}(\ve,\pe)$}
If we are interested in the value of $J_{1,2,3}$ at $p=\pe$ then
the range of momentum integration is the sphere of convergence of the disorder structure factor $g$, namely the sphere  of
radius $p_0$ with the center at the external momentum $\bq=\pe\bn$.
There are two  distinctive contributions. The first one comes from
the integration inside the small sphere: $|\bp-\bn\pe|\le 1/(\tau
v)$, where the integrand is peaked due to the proximity of the poles of $G^AG^R$; the second one comes from a thick shell $
|1/(\tau v_{\textrm{f}})|\lesssim |p-\pe\bn|\lesssim p_0 $.
Alternatively, one can think of the first contribution as the one
coming from the singularities of $G^RG^A$ and the second of from
the singularities of $g_{\bn\pe-\bp}$. Simple estimates give
\begin{gather}
  \label{estimates}
  \begin{split}
  \int\limits_{|\bp-\bn\pe|\lesssim \frac{1}{\tau v}}
  \hskip-3mm d\bp G^A(\bp) J_\alpha G^R(\bp)  g_{\bn\pe-\bp} \sim
  \tau\ve\frac{\pe}{v^2} J_\alpha(\pe) g_\pe,\\
  \int\limits_{\frac{1}{\tau v}\lesssim p\lesssim
  |\pe\bn+\bn^\prime p_0|}\hskip-6mm d\bp G^A(\bp) J_\alpha G^R(\bp)
   g_{\bn\pe-\bp} \sim
  \frac{p^3_0}{\ve^2}g_{p_0}
 \end{split}
  \end{gather}
Comparing the two contributions, we see that if the disorder cut-off $p_0$ is not too high, the contribution from the thick shell is
$1/(\tau\ve)\times(p_0/\pe)^3$ smaller than the one from the small
sphere due to the $G^RG^A$product. Therefore, one can indeed discard the contribution from the shell $ |1/(\tau v)|\lesssim |p-\pe\bn|\lesssim|p_0|$ (or equivalently, the contribution from the poles of the potential
$g_{\bn\pe-\bp}$). Thus, in the leading order of the $1/(\tau\ve)$-expansion, one can take into account only the poles of the $G^RG^A$ term.

\subsection{Vertex $\mathbf{J}^{ARR}(\ve,p)$ at
$p\rightarrow\infty$}

The same logic of estimates holds for the large $p$-behavior of the
vertex. We now plug ansatz~\eqref{ansatz-iso} into the Dyson
equation~\eqref{vertex1a} and employ the estimates~\eqref{estimates},
modified for the external momentum $q\gg\pe,\ p_0$. The computation is simple though tedious. Solving the system of self-consistent
equations, one arrives at the following asymptotics

\begin{gather}
   \begin{split}
    J_1(q)&=1+\beta_1g_q\frac{\pe^2}{p_0^2}J,\\
    J_2(q)&=\beta_2\frac{q}{\pe}p_0\de_qg_q
    \frac{\pe^2}{p_0^2}J,\\
    J_3(q)&=\beta_3g_q\frac{\pe^2}{p_0^2}J\, ,
   \end{split}
 \end{gather}
where $J\equiv J_1(\pe)+J_2(\pe)+J_3(\pe)$ and $\beta_{1,2,3}$ are constants of the order of unity. We see that the
estimate for the vertex in the isotropic case can be rewritten as
 \begin{gather}
   \label{final-estimate}
   \mathbf{J}(\bq)-\bsigma\sim\mathbf{r}g(q)\,,
\end{gather}
where $\mathbf{r}$ is some restricted vector function of the momentum $\bq$.
Therefore, the renormalized current vertex  $ \mathbf{J}(\bq)$ approaches the bare one $\bm{\sigma}$ at large momenta $q$ with the
rate proportional to the structure factor $g(q)$.

\subsection{\textit{On shell} Dyson equation}
Now we are able to write down the Dyson equation in a somewhat simplified form performing  integration over the absolute value of the momentum. As was argued previously, the most important contribution comes from the peak in the $G^RG^A$ product at $p^\prime=\pe$. Therefore, we put $p^\prime$ in the argument of $\mathbf{J}(\ve,p^\prime\bn^\prime)$ and
$g^\prime(\bp-p^\prime\bn^\prime,\bp_0)$ inside the integral equation~\eqref{vertex1a} equal to $\pe$. Moreover, we take into account the fact that in all the limiting cases considered (isotropy, anisotropy with short and long-range disorder) the Green's function assumes a simplified form~\eqref{green-2} rather than~\eqref{green-exact} (see Appendix F).
Changing $p^2=\zeta,\ \ dp=v_\parallel d\zeta/2\ve$, and changing the lower limit of integration to $-\infty$, we perform the integration
\begin{gather}
   \int\limits_0^\infty\!\frac{p^2dp}{\Big[\Big(\ve\!+\!
   \frac{i}{2\tau}\Big)^2\!-\!v_\parallel^2\Big(p-\frac{i}{2\tau_1}\Big)^2\Big]
       \Big[\Big(\ve\!-\!\frac{i}{2\tau}\Big)^2-v_\parallel^2\Big(p+
       \frac{i}{2\tau_1}\Big)^2\Big]}\notag\\
       =\frac{\pi}{2v_\parallel^3}\frac{1}{\frac{1}{\tau}+\frac{1}{\tau_1}}.
 \end{gather}
This way we arrive at the Dyson equation~\eqref{dyson-sphere} and~\eqref{dyson-sphere1} for the modified vertex.

\section{Estimate of the \textit{regular} term in conductivity}
Thanks to the factor $\ve^2-p^2$ in the nominator of~\eqref{est1} we
can substitute
\begin{gather}
   (p^2-\ve^2)G^A(\ve,\bp)\approx (\ve+\bp\bsigma)\Big[1+
   \mathcal{O}\Big(\frac{1}{\ve\tau}\Big)\Big]\, ,
\end{gather} and the contour of integration is no longer  squeezed
between the proximate poles of the product $G^RG^A$. The path
therefore can be deformed in any suitable way. Estimating at
$p\sim\hbox{max}\{\pe,p_0\},\ \ G^RG^A\sim1/\ve^2$, we obtain
\begin{gather}
 \hbox{I}\sim \delta_{\alpha\beta}
 |\max\{p_\ve|\mathbf{J}(\ve,\pe)-\bsigma|,p_0|\mathbf{J}
 (\ve,p_0)-\bsigma|\}\frac{1}{v^2}
  \end{gather}
On the other hand, the convergent expression II is completely
determined by the behavior of $J_\alpha^{ARR}(\ve,p)$ in the
vicinity of the peak and one can substitute $J_\alpha(\ve,p)$ with
$J_\alpha(\ef,\pf)$. The integral is then estimated as
\begin{gather} \label{II}
   \hbox{II}\sim\delta_{\alpha\beta}\big|\mathbf{J}(\ve,\pe)-\bsigma\big|
  \frac{\ve^2\tau}{v^3}.
\end{gather}
Combining Eqs.~\eqref{II} and \eqref{final-estimate}, we obtain relation~\eqref{comp}.

\section{Conductivity, short-range potential}
In the $\delta$-correlated (momentum-independent) case, the vertex functions $\Psi$ and $\Phi$ are angle-independent and, as follows from Eq.~\eqref{dyson-sphere}
\begin{gather}
   \Phi(\bn\bn_0)\equiv\Phi_0=\frac{3\tau}{2},\ \ \Psi(\bn\bn_0)=0.
\end{gather}
In the case of small momentum-dependent correction to form factor \eqref{form1}, these also acquire small angle-dependent corrections
\begin{gather}
   \Phi(\bn\bn_0)=\Phi_0+\delta\Phi(\bn\bn_0),\ \ \Psi(\bn\bn_0)=\delta\Phi(\bn\bn_0)
\end{gather}
of the order of $(\pe/p_0)^2$.
Thus, the formula for the conductivity (see Eq.~\eqref{cond-fin}) becomes
\begin{gather}
  \begin{split}
   \label{cond-fin1}
    &\sigma_{\alpha\beta}=\sigma\\
    &\times\Big(\delta_{\alpha\beta}+
      \frac{\delta\Phi_0-\delta\Phi_2}{4\tau}\delta_{\alpha\beta}
      +\frac{3\delta\Phi_2-\delta\Phi_0+2\delta\Psi_1}
      {4\tau}n_{0\alpha}n_{0\beta}\Big)\, ,
   \end{split}
\end{gather}
where $(\delta\Psi_n,\delta\Phi_n)=\int_{-1}^1(\delta\Psi,\delta\Phi)x^n\,dx$.
Now $\delta\Phi_n$ and $\delta\Psi_n$ are easily extracted perturbatively from the Dyson equation (see the next Appendix for details).
The corrections can be written as
\begin{gather}
   \begin{split}
   \delta\sigma_\bot=\xi^2\sigma\frac{\delta\Phi_0-
   \delta\Phi_2}{4\tau},\\
   \delta\sigma_\parallel=\sigma\frac{\delta\Phi_2+
   \delta\Psi_1}{2\tau}\, .
      \end{split}
\end{gather}
We thus obtain the conductivity as a function of the energy
\begin{gather}
  \label{cond-short}
   \begin{split}
   \delta\sigma_\bot&=-\sigma\frac{\xi^2}{5}\frac{\pe^2}
   {p_0^2}\frac{\de g}{g}(1+4\xi^{-2}),\\
   \delta\sigma_\parallel&=-\sigma\frac{\xi^2}{15}
   \frac{\pe^2}{p_0^2}\frac{\de g}{g}(7+8\xi^{-2})\,.
   \end{split}
\end{gather}
Plugging these into~\eqref{cond-temp}, we determine the temperature dependence of the conductivity~\eqref{cond-short0}.

\section{Solution of the anisotropic Dyson equation}
\subsection{Short-range potential}

We now need to compute the anisotropic corrections to the spherically-symmetric self-energy. From the structure of~\eqref{im-self2}, we see that we need to compute the following averages
\begin{gather}
  \begin{split}
  &\frac{n_{\text{imp}}\ve^2}{4v_0^3\pi\xi^2}\int\frac{d\Omega^\prime}
  {4\pi}g^\prime_{\pe(\bn-\bn^\prime),\bn_0}\bn\\
  &=
  -\frac{\de g(0)\pe^2}{3\tau p_0^2}\big[(1-
  \xi^{-2})(\bn\bn_0)\bn_0+\xi^{-2}\bn\big]\, ,\\
  &\frac{n_{\text{imp}}\ve^2}{4v_0^3\pi\xi^2}\int\frac{d\Omega^\prime}{4\pi}  g^\prime_{\pe(\bn-\bn^\prime),\bn_0}n_\alpha n_\beta=
   \frac{1}{6\tau}\\
   &\times\bigg(\delta_{\alpha\beta}\bigg[
   1+\frac{\de g(0)\pe^2}{p_0^2}\bigg\{(1-
   \xi^{-2})(\bn\bn_0)^2+\frac{12+
   9\xi^{-2}}{5}\bigg\}\bigg]\\
   &+\frac{2\de g(0)\pe^2}{5p_0^2}(1-
   \xi^{-2})(\bn\bn_0)n_{0\alpha}n_{0\beta}\bigg)\, .
  \end{split}
\end{gather}
Therefore, we see that the self-energy retains its isotropic form
with the corrections to the scattering times:
\begin{gather}
   \delta\frac{1}{\tau}=
 \frac{1}{\tau}\frac{\de g(0)p_\ve^2}{p_0^2}\bigg((1-
 \xi^{-2})\Big[(\bn\bn_0)^2+\frac{1}{3}\Big]+2\xi^{-2}\bigg),\\
   \delta\frac{1}{\tau_{\textrm{tr}}}=\frac{1}{\tau}
   \frac{\de g(0)\pe^2}{p_0^2}\bigg(\frac{5}{3}(1-
   \xi^{-2})(\bn\bn_0)^2+\frac{1+
   7\xi^{-2}}{3}\bigg).\notag
\end{gather}
Following the same steps as in the isotropic case, we arrive at the following equations for the vertex functions
\begin{gather}
 \label{self}
 \begin{split}
 \delta\Phi(x)=-\frac{2\tau}{5}\de g(0)\frac{\pe^2}{p_0^2}
 (1+4\xi^{-2})+\frac{\delta\Phi_0-
 \delta\Phi_2}{4},\\
 \delta\Psi(x)=-\frac{8\tau}{15}\de g(0)\frac{\pe^2}{p_0^2}
 (1-\xi^{-2})x+\frac{\delta\Psi_0+
 \delta\Psi_1}{2}x\, .
 \end{split}
\end{gather}
Solving Eqs.~\eqref{self} in a self-consistent way, we find
\begin{gather}
  \begin{split}
   \delta\Phi_0=-\frac{6}{5}\tau(1+
   4\xi^{-2})\de g(0)\frac{\pe^2}{p_0^2},\\
   \delta\Phi_2=-\frac{2}{5}\tau(1+
   4\xi^{-2})\de g(0)\frac{\pe^2}{p_0^2},\\
   \delta\Psi_1=-\frac{8}{15}\tau(1-
   \xi^{-2})\de g(0)\frac{\pe^2}{p_0^2}.
   \end{split}
\end{gather}
From the last equations, we easily recover Eqs.~\eqref{cond-short}.

\subsection{Long-range disorder}
Expanding in the scattering angle, we have
 \begin{gather}
   \label{expansion}
  \begin{split}
 (\bn-\bn^\prime)\bn_0&=(1-\cos\theta^\prime)\cos\theta+
 \sin\theta^\prime\sin\theta\cos\vf\\
 &\approx \theta^\prime\sin\theta\cos\vf,\\
 (\bn-\bn^\prime)^2&=2-2\cos\theta\approx\theta^{\prime2},\\
 do^\prime&=\sin\theta^\prime d\theta^\prime d\vf\approx
 \theta^\prime d\theta^\prime d\vf\, .
 \end{split}
\end{gather}
Next, we extend the upper limit of integration over $\theta^\prime$
to $+\infty$, and we obtain
 \begin{gather}
   \label{self-long}
   \begin{split}
   &\hbox{Im}\,\Sigma(\ve,\pe\bn)=-\frac{n_{\text{imp}}\ve^2u_0^2}
   {4\pi v_\parallel^3p_0^6\xi^2}\int\limits_0^{2\pi}\frac{d\vf}{2\pi}
   \begin{pmatrix}
      1+\sigma_x\theta^\prime\cos\vf\\
      1+\sigma_y\theta^\prime\sin\vf\\
      1-\sigma_z\frac{\theta^{\prime2}}{2}
   \end{pmatrix}\\
   &\times\int_0^\infty\frac{\theta^\prime d\theta^\prime}{2}
  g\left(\frac{\pe^2\theta^{\prime2}}{p_0^2}[(1-\xi^{-2})
  \sin^2\theta\cos^2\vf+\xi^{-2}]\right)\\
  &=
  -\frac{1}{2\tau(\bn\bn_0)}+\frac{\bsigma\bn}{2}\Big(\frac{1}
  {\tau(\bn\bn_0)}-\frac{1}{\tau_{\textrm{tr}}(\bn\bn_0)}\Big)\, .
  \end{split}
\end{gather}
In addition to~\eqref{sc}, we will need one more scattering time
\begin{gather}
  \begin{split}
        \frac{1}{\tau_{\textrm{tr}_0}(\ve)}&=\frac{n_{\text{imp}}\ve^2u_0^2}
        {2\pi \xi^2 v_\parallel^3p_0^6\xi^2}\int\frac{d\Omega}{4\pi}g^\prime_{\pe(\bn-\bn^
        \prime),\bn_0}(1-\cos\theta_{\bn\bn^\prime})\cos^2\vf\\
  \end{split}
 \end{gather}
(see Fig.~\ref{weyl3} for the definition of the angle $\vf$). To obtain the explicit formulae for all scattering times, we expand the argument of
$g^\prime_{\pe(\bn-\bn^\prime),\bn_0}$ and integration domain
$d\Omega$.

Introducing the notation
\begin{gather}
  \begin{split}
    \frac{1}{\tau_0(\ve)}&=\kappa^2\frac{\ve^2n_{\text{imp}}g_0 u_0^2}
    {8\pi v_\parallel^3p_0^6},\\
    \Delta(\theta)\equiv\Delta(\bn\bn_0)&=[(\xi^2-1)
    \sin^2\theta+1]^{1/2}\,,
  \end{split}
 \end{gather}
where  $g_0$ is defined after Eq.~\eqref{dyson-fin1},
we obtain results for the cross-sections
\begin{gather}
   \begin{split}
   \frac{1}{\tau(\bn\bn_0)}&=\frac{1}{\tau_0}
   \frac{1}{\Delta(\bn\bn_0)}\, , \\
   \frac{1}{\tau_{\textrm{tr}_0}}&=\frac{\kappa^2\xi^2}
   {4}\frac{g_1}{g_0\tau_0}\frac{1}{\Delta^3(\bn\bn_0)}\, ,\\
   \frac{1}{\tau_{\textrm{tr}}(\bn\bn_0)}&=\frac{1}
   {\tau_{\textrm{tr}_0}(\bn\bn_0)}[\Delta^2(\bn\bn_0)+1]\, .\\
  \end{split}
 \end{gather}
(Here $g_1$ is defined after Eq.~\eqref{tau-tr2}).
As one can see from the expression for $\text{Im}\,\Sigma$~\eqref{self-long}, the Green's function retains its simple structure~\eqref{green-2}. One needs to pay particular attention to the fact that the scattering time is now a function of the direction of the incoming particle with respect to the anisotropy axis. Now we plug the Green's functions~\eqref{green-2} and ansatz~\eqref{ans1} into the Dyson equation~\eqref{dyson-sphere1}. The most obvious step in solving the Dyson equation~\eqref{dyson-sphere1} is to take the zero order in the $\kappa$-expansion and to set $\bn^\prime=\bn$ in the
peaked integrand. One immediately arrives at an algebraic equation
for all the constants. Despite the fact that one is able to
immediately obtain some of the vertices: $I_1=1,\ I_5=I_7=0$, we
also obtain a meaningless equation $I_2=I_3=1+I_2$ and the vertices
$I_4,\ I_6$ remain undefined. This is a signature of the fact that
one needs to take the next term in the $\kappa$-expansion.
The $\kappa$ expansion in turn leads to the emergence of the derivatives of functions $\mathbf{I}$ in the structure of the equation.

We use the commutation relation and find
$(1+\bn\bsigma)\bsigma(1+\bn\bsigma)=2\bn(1+\bn\bsigma)$.
\begin{gather}
  \begin{split}
  &(1+\bn^\prime\bsigma)\mathbf{I}(1+\bn^\prime\bsigma)\\
  &=2(I_1+I_2+I_3+I_5\bn^\prime\bn_0)\bn(1+\bn^\prime\bsigma)\\
  &+2(I_4++I_6+I_7\bn^\prime\bn_0)\bn_0(1+\bn^\prime\bsigma).
 \end{split}
 \end{gather}
 Whenever necessary, we expand
 \begin{gather}
 \begin{split}
&\bn^\prime=\bn+\delta\bn,\\
&I(\bn^\prime\bn_0)=I(\bn\bn_0+\delta\bn\bn_0)\\
&=I(\bn\bn_0)+\bn_0\delta\bn\de_{\bn\bn_0} I(\bn\bn_0)+
\frac{(\bn_0\delta\bn)^2}{2}\de^2_{\bn\bn_0} I(\bn\bn_0)+...
 \end{split}
 \end{gather}
 Next, we perform the integration over angles
\begin{gather}
   \begin{split}
   &\frac{n_{\text{imp}}u_0^2\ve}{4v_\parallel^3p_0^6\xi^2\pi}
   \int\frac{d\Omega^\prime}{4\pi}g^\prime_{\pe(\bn-bn^\prime),\bn_0}\delta
   \bn=-\frac{\bn}{2\tau_{\textrm{tr}}}\, ,\\
  &\frac{n_{\text{imp}}u_0^2\ve^2}{4v_\parallel^3p_0^6\xi^2\pi}\int
  \frac{d\Omega^\prime}{4\pi}g^\prime_{\pe(\bn-bn^\prime),\bn_0}\delta n_\alpha\delta n_\beta\\
  &=\frac{1}
  {\tau_{\textrm{tr}_0}}
   \Big[\Delta^2\delta_{\alpha\beta}-(\xi^2-1)n_
   {0\alpha}n_{0\beta}\\
   &-\xi^2n_\alpha n_\beta+
   (\xi^2-1)(\bn\bn_0)
   (n_{0\alpha}n_\beta+n_{\alpha}n_{0\beta})\Big]\, .
  \end{split}
  \end{gather}
The Dyson equation is split into seven differential equations. In terms of $\Phi(x)$ and $\Psi(x)$, we rewrite it as follows
\begin{widetext}
\begin{gather}
   \label{dyson-fin}
   \begin{split}
        I_1\left(\frac{2}{\tau}-\frac{1}{\tau_{\textrm{tr}}}\right)&
        =1+\frac{2\Delta^2}{\tau_{\textrm{tr}}}\Phi,\\
        I_2\left(\frac{2}{\tau}-\frac{1}{\tau_{\textrm{tr}}}\right)&
        =\Phi\left(\frac{1}{\tau}-\frac{1}{\tau_{\textrm{tr}}}\right)-
        x\Phi^\prime\left(\frac{2}{\tau_{\textrm{tr}_0}}+\frac{1}
        {\tau_{\textrm{tr}}}\right)+
        \frac{1}{\tau_{\textrm{tr}_0}}\Phi^{\prime\prime}(1-x^2)
        +I_5\frac{x}{\tau}, \\
        I_3\left(\frac{2}{\tau}-\frac{1}{\tau_{\textrm{tr}}}\right)&=
        \Phi\left(\frac{1}{\tau}-\frac{2}{\tau_{\textrm{tr}}}-
        \frac{2\xi^2}{\tau_{\textrm{tr}_0}}\right)-
        x\Phi^\prime\left(\frac{4}{\tau_{\textrm{tr}_0}}+\frac{1}
        {\tau_{\textrm{tr}}}\right)+ \frac{1}{\tau_{\textrm{tr}_0}}
        \Phi^{\prime\prime}(1-x^2)+I_5\frac{x}{\tau}, \\
       I_{4}\left(\frac{2}{\tau}-\frac{1}{\tau_{\textrm{tr}}}\right)&
       =\frac{2}{\tau_{\textrm{tr}_0}}\left(\Phi(\xi^2-1)x+\Phi^\prime\right)
      +\Psi\left(\frac{1}{\tau}-\frac{1}{\tau_{\textrm{tr}}}
      \right)-x\Psi^\prime\left(\frac{2}{\tau_{\textrm{tr}_0}}+
      \frac{1}{\tau_{\textrm{tr}}}\right)
      +\frac{1}{\tau_{\textrm{tr}_0}}\Psi^{\prime\prime}(1-x^2), \\
       I_{5}\left(\frac{2}{\tau}-\frac{1}{\tau_{\textrm{tr}}}\right)&
       =\frac{2}{\tau_{\textrm{tr}_0}}\left(\Phi(\xi^2-1)x+\Phi^\prime\right), \\
        I_6\left(\frac{2}{\tau}-\frac{1}{\tau_{\textrm{tr}}}\right)&
        =\frac{2}{\tau_{\textrm{tr}_0}}\Phi^\prime
   +\frac{1}{\tau}\Psi-\frac{x}{\tau_{\textrm{tr}}}\Psi^\prime+
   \frac{1}{\tau_{\textrm{tr}_0}}\Psi^{\prime\prime}(1-x^2), \\
       \left(\frac{2}{\tau}-\frac{1}{\tau_{\textrm{tr}}}\right)I_7&
       =-\frac{2}{\tau_{\textrm{tr}_0}}\Phi(\xi^2-1)+
       \frac{2}{\tau_{\textrm{tr}_0}}\Psi^\prime\, .
   \end{split}
 \end{gather}
\end{widetext}
Summing up the equations, we obtain~\eqref{dyson-fin1}.

\end{document}